\documentclass[american,aps,pra,reprint,groupedaddress,twocolumn,showpacs]{revtex4-2} 
\usepackage[unicode=true,pdfusetitle, bookmarks=true,bookmarksnumbered=false,bookmarksopen=false, breaklinks=false,pdfborder={0 0 0},backref=false,colorlinks=false] {hyperref}
\hypersetup{ colorlinks,linkcolor=myurlcolor,citecolor=myurlcolor,urlcolor=myurlcolor}
\usepackage{braket,colortbl,amsthm,amsmath,amssymb,txfonts}
\definecolor{myurlcolor}{rgb}{0,0,0.7}
\usepackage{graphics,graphicx}
\usepackage{color}
\usepackage{xcolor}
\usepackage{colortbl}
\usepackage{subcaption}
\usepackage{bbm}

\usepackage{dsfont}
\usepackage{float}
\usepackage{hyperref}
\usepackage[font=small,labelfont=bf,
justification= centerlast,
format=hang]{caption}

\usepackage{ragged2e}
\usepackage{color}
\usepackage{braket}
\usepackage{physics}
\usepackage{tikz}
\usepackage{color}
\usepackage{braket}
\usepackage{physics}
\usepackage{tikz}
\usetikzlibrary{shapes.geometric, arrows}
\usepackage{latexsym}
\usepackage{bm}
\usepackage{graphics,epstopdf}
\usepackage{color}
\usepackage[usenames,dvipsnames,svgnames]{pstricks}
\usepackage{epsfig}
\usepackage{pst-grad} 
\usepackage{pst-plot} 
\usepackage[mathscr]{euscript}

\begin{document}
\title{Linear optics and the problem of Bell-like state discrimination}
\author{Jatin Ghai$^{1,2}$}
\email{jghai@imsc.res.in}
\author{Sibasish Ghosh$^{1,2}$}
\email{sibasish@imsc.res.in}
\affiliation{$^1$Optics and Quantum Information Group, The Institute of Mathematical Sciences, C. I. T. Campus, Taramani, Chennai 600113, India}
\affiliation{$^2$Homi Bhabha National Institute, Training School Complex, Anushaktinagar, Mumbai 400094, India}
\begin{abstract}
 A linear optics-based scheme to implement various quantum information processing tasks is of paramount importance due to its ease of implementation and low noise. Many information-theoretic tasks depend on the successful discrimination of Bell states. A no-go theorem states that it is not possible to perfectly discriminate among the four Bell states when the measurement apparatus is restricted to linear optical elements. The success probability is only $50\%$. Using extra resources such as hyperentanglement, ancillary entanglement, and even a minimum amount of non-linearity, complete Bell-state discrimination can be achieved. The success probability for Bell-like state discrimination is only $25\%$. Bell-like states are non-maximally entangled two-qubit states that resemble Bell states in structure but have unequal Schmidt coefficients. We find that this can be boosted to $50\%$ using hyperentanglement, where extra ancillary degrees of freedom (DOF) of photons, such as polarization, momentum, or OAM, are also entangled along with the system DOF. This contrasts with Bell-state discrimination, where $100\%$ can be achieved. Furthermore, we find that using correlations in the photons' arrival times, all four Bell states can be distinguished with $100\%$ success probability, while for Bell-like state discrimination, it increases only to $37.5\%$ from $25\%$. We observe a similar contrast when using ancillary entangled photons. While the success probability for discriminating all four Bell states increases as $1-\frac{1}{2^N}$, where $N$ is the number of ancillary photon pairs, for Bell-like states, it depends on the state parameters and can be less than $25\%$ in some cases. Also, adding further ancillary photons decreases the success probability. We then show that using non-linear gadgets, namely SFG, $100\%$ success probability can be achieved even for Bell-like state discrimination.
\end{abstract}
\maketitle
\section{Introduction}
Linear optics is one of the leading platforms for quantum information processing, offering a promising approach to both quantum computation\cite{RevModPhys.79.135} and quantum communication\cite{azuma2015all}. Their appeal stems from the fact that photons, which serve as the information carriers, are highly robust against environmental noise and possess long coherence times, making them well suited for implementing quantum information processing tasks. Moreover, the Knill--Laflamme--Milburn (KLM) scheme demonstrated that universal quantum computation is, in principle, possible using only linear-optical elements together with measurement-induced nonlinearities\cite{knill2001scheme,takeda2019toward,carolan2015universal}. Any N-port unitary operation can be implemented using an array of $\frac{N(N-1)}{2}$ beam splitters\cite{reck1994experimental,PhysRevX.8.021017}. This number grows only quadratically with N, thus giving a realistic hope for implementing various quantum operations using linear optical elements. But there are fundamental limitations on what kind of operations are possible to implement deterministically if we stick to linear optics only\cite{PhysRevA.69.012302,PhysRevA.73.062320,PhysRevA.108.049901,MOYANOFERNANDEZ2017237}. A very simple but important example of the inadequacy of the linear optical-based schemes is that it is unable to distinguish among maximally entangled Bell states if no ancillary photons and resources are allowed\cite{lutkenhaus1999bell}. By relaxing the constraints of perfect Bell state discrimination by allowing it to fail with some probability, it can be proved that the maximum success probability of this scheme is $50\%$ given that there are zero auxiliary photons and no conditional measurements\cite{calsamiglia2001maximum}. Thus, linear optics-based schemes are restrictive in that sense. But various information-theoretic tasks namely quantum teleportation\cite{bouwmeester1997experimental}, entanglement swapping\cite{zhang2017simultaneous,schmid2009quantum}, super-dense coding\cite{barreiro2008beating,williams2017superdense}, etc. depend on the successful discrimination of Bell states. Hence, it is important to look for ways of increasing the probability of successful Bell state discrimination from $50\%$. As it turns out, it can be increased by using extra ancillary photons\cite{PhysRevA.84.042331}, hyperentanglement\cite{PhysRevA.68.042313,PhysRevLett.118.050501,barreiro2008beating}, or some amount of non-linearity\cite{PhysRevLett.86.1370,PhysRevLett.110.260501}. All these methods can boost the success probability to $100\%$.

A more general version of this problem is distinguishing among the Bell-like states unambiguously. On the same lines as the treatment in \cite{calsamiglia2001maximum} it was shown in \cite{PhysRevA.106.023706} that the optimal success probability for discrimination of Bell-like states without using any ancillary photons and conditional measurements is $25\%$. As Bell-like states are asymmetric in structure, this optimal success probability can be achieved by introducing an unbalanced beam splitter in the path of appropriate modes of the Bell-like state. This leads to cancellations of some unique detection signatures for a pair of these states, which facilitates their discrimination. For further details, we refer the reader to \cite{PhysRevA.106.023706}. In practical scenarios, one might not be provided with Bell states. Instead, a weaker resource, namely non-maximally entangled states, might be present.  Thus, it is important to distinguish among non-maximally entangled states. Also, it has recently been shown that, in order to distribute the state with the highest teleportation fidelity through entanglement swapping, sometimes non-maximally entangled measurements outperform the Bell measurements\cite{quantteleport,ghosal2024repeater}. Thus, it is worthwhile to explore the distinguishability problem of Bell-like states.

In this work, we try to devise methods to enhance the success probability of unambiguous Bell-like state discrimination using extra resources on the same lines as done in the case of Bell-state discrimination and compare the two. We, for the time being, are not concerned about the specific preparation procedure (that is, using linear optics or otherwise) of these states. We should also mention here that throughout the work we are using the dual-rail encoding (see Sec. II below for its definition) for the photons, and while distinguishing the two-photon entangled states, no restriction is going to be imposed on the accessibility of both the photons: they can be either in one lab or in spatially separated labs. The brief outline of the paper is as follows:

In Sec. (\ref{sec2}) ancillary entanglement in the polarization basis of photons is used as a resource in the form of hyperentanglement, while system qubits are represented by their momentum DOF. In Sec. (\ref{sec3}) and (\ref{sec4}), a similar thing is done for ancillary entanglement in the momentum DOF and OAM degrees of freedom, respectively, with system qubits in the polarisation basis in both cases. In Sec. (\ref{sec5}), correlations in time are used to assist in distinguishing Bell-like states. Sec.  (\ref{sec6}) includes the effect of ancillary entangled physical qubits on the success probability of unambiguous discrimination of Bell-like states while  Sec. (\ref{sec7}) deals with the effect of using a non-linear resource called SFG in the context of distinguishing Bell-like states. Finally, we conclude and try to give some future directions in Sec. (\ref{sec8}). All results are compared with the corresponding scenario of the Bell-state discrimination. 

\section{Using Hyperentanglement for Bell-like state discrimination}
We are given the following four mutually orthogonal Bell-like states:
 \begin{align}
            &\vert \psi_1\rangle=(\sin(\theta_1)\hat{a}^{\dag}_1\hat{a}^{\dag}_2+\cos(\theta_1)\hat{b}^{\dag}_1\hat{b}^{\dag}_2)\ket{0},\label{1}\\
            &\vert \psi_2\rangle=(\cos(\theta_1)\hat{a}^{\dag}_1\hat{a}^{\dag}_2-\sin(\theta_1)\hat{b}^{\dag}_1\hat{b}^{\dag}_2)\ket{0},\label{2}\\
            &\vert \psi_3\rangle=(\sin(\theta_2)\hat{a}^{\dag}_1\hat{b}^{\dag}_2+\cos(\theta_2)\hat{b}^{\dag}_1\hat{a}^{\dag}_2)\ket{0},\label{3}\\
            &\vert \psi_4\rangle=(\cos(\theta_2)\hat{a}^{\dag}_1\hat{b}^{\dag}_2-\sin(\theta_2)\hat{b}^{\dag}_1\hat{a}^{\dag}_2)\ket{0},\label{4}
        \end{align}
        where $\{\hat{a}^{\dag}_{1(2)}, \hat{b}^{\dag}_{1(2)}\}$ are the creation operators for the first (second) photon. Throughout this paper, we will use the dual-rail encoding. Dual-rail representation is one of the most common encoding schemes for photonic qubits. In this encoding, the logical qubit is represented by the presence of a single photon distributed across two orthogonal optical modes  (like horizontal and vertical polarizations, +1 OAM and -1 OAM values, etc.). The logical basis states are defined as
\begin{align}
|0_L\rangle&=|10\rangle=\ket{1}\otimes\ket{0},\\
|1_L\rangle&=|01\rangle=\ket{0}\otimes\ket{1},
\end{align}
where the numbers denote the photon occupation of the two optical modes. Thus, written as $|0_L\rangle$ and $|1_L\rangle$, both of these logical qubit states represent a single photon occupying one of two optical modes. The four states above are therefore two-photon states, with one photon encoding each logical qubit. Here, $\hat{a}^{\dag}_{1(2)}$ corresponds to the creation operator for the first (second) photon in state $|0_L\rangle_{1(2)}$, while $\hat{b}^{\dag}_{1(2)}$ corresponds to the first (second) photon in state $|1_L\rangle_{1(2)}$. The four Bell-like states can be expressed in terms of these logical qubits as
\begin{align}
    |\psi_1\rangle &= \sin(\theta_1)|0_L\rangle_1|0_L\rangle_2
+\cos(\theta_1)|1_L\rangle_1|1_L\rangle_2,\\
|\psi_2\rangle &= \cos(\theta_1)|0_L\rangle_1|0_L\rangle_2
-\sin(\theta_1)|1_L\rangle_1|1_L\rangle_2,\\
|\psi_3\rangle &= \sin(\theta_2)|0_L\rangle_1|1_L\rangle_2
+\cos(\theta_2)|1_L\rangle_1|0_L\rangle_2,\\
|\psi_4\rangle &= \cos(\theta_2)|0_L\rangle_1|1_L\rangle_2
-\sin(\theta_2)|1_L\rangle_1|0_L\rangle_2.
\end{align}

Dual-rail encoding provides a convenient framework for photonic quantum information processing, wherein a qubit is encoded in the occupation of two orthogonal optical modes. Since each logical basis state contains exactly one photon, arbitrary single-qubit operations can be implemented using passive linear-optical elements, such as beam splitters and phase shifters. Furthermore, dual-rail encoding enables the preparation of arbitrary single-qubit states through linear-optical interference, while measurements can be performed by detecting the photon in either of the two modes. The Bell-like states can be produced experimentally using the process of SPDC\cite{PhysRevA.68.042313}. Our job is to distinguish them unambiguously using passive linear optical elements like beam splitters, wave plates, etc., and compare it to the corresponding case of Bell state discrimination. It has been shown in \cite{PhysRevA.106.023706} that by confining ourselves to the regime of linear optics, we can distinguish only between two Bell-like states $\vert \psi_1\rangle$ and $\vert \psi_2\rangle$ (or $\vert \psi_3\rangle$ and $\vert \psi_4\rangle$) with a success probability of $25\%$. This is half compared to the optimal success probability for Bell states\cite{calsamiglia2001maximum}, which is $50\%$ under the same settings. For Bell states, it is known that by using an ancillary pair of entangled photons, this success probability can be boosted to $75\%$\cite{PhysRevA.84.042331}. Besides this, using extra entanglement in other degrees of freedom of the same photon pair, like momentum \cite{PhysRevA.68.042313}, time-bin \cite{PhysRevA.58.R2623, PhysRevLett.118.050501}, orbital angular momentum\cite{barreiro2008beating} etc., it is known that this can be boosted to $100\%$. This ancillary enatanglement in other degree of freedom of photon is known as hyperentanglement.
\subsection{Ancillary entanglement in polarisation degrees of freedom}\label{sec2}
   In this approach, momentum degrees of freedom of photons are chosen as system degrees of freedom, while polarisation degrees of freedom of these photons are used for ancillary entanglement. We will follow the procedure described in \cite{PhysRevA.68.042313} very closely. The joint states entangled in both momentum and polarization degrees of freedom are written as 
\begin{align}
            &\vert \Theta_1\rangle=(\sin(\theta)\hat{a}^{\dag}_1\hat{a}^{\dag}_2+\cos(\theta)\hat{b}^{\dag}_1\hat{b}^{\dag}_2)\otimes\frac{(\hat{h}^{\dag}_1\hat{v}^{\dag}_2+\hat{v}^{\dag}_1\hat{h}^{\dag}_2)}{\sqrt{2}}\ket{0},\label{5}\\  
            &\vert \Theta_2\rangle=(\cos(\theta)\hat{a}^{\dag}_1\hat{a}^{\dag}_2-\sin(\theta)\hat{b}^{\dag}_1\hat{b}^{\dag}_2)\otimes\frac{(\hat{h}^{\dag}_1\hat{v}^{\dag}_2+\hat{v}^{\dag}_1\hat{h}^{\dag}_2)}{\sqrt{2}}\ket{0},\\
            &\vert \Theta_3\rangle=(\sin(\theta)\hat{a}^{\dag}_1\hat{b}^{\dag}_2+\cos(\theta)\hat{b}^{\dag}_1\hat{a}^{\dag}_2)\otimes\frac{(\hat{h}^{\dag}_1\hat{v}^{\dag}_2+\hat{v}^{\dag}_1\hat{h}^{\dag}_2)}{\sqrt{2}}\ket{0},\\          
            &\vert \Theta_4\rangle=(\cos(\theta)\hat{a}^{\dag}_1\hat{b}^{\dag}_2-\sin(\theta)\hat{b}^{\dag}_1\hat{a}^{\dag}_2)\otimes\frac{(\hat{h}^{\dag}_1\hat{v}^{\dag}_2+\hat{v}^{\dag}_1\hat{h}^{\dag}_2)}{\sqrt{2}}\ket{0}.\label{8}
\end{align}

\begin{figure}
     \centering
     \begin{subfigure}[b]{\linewidth}
         \centering
         \includegraphics[width=0.95\linewidth]{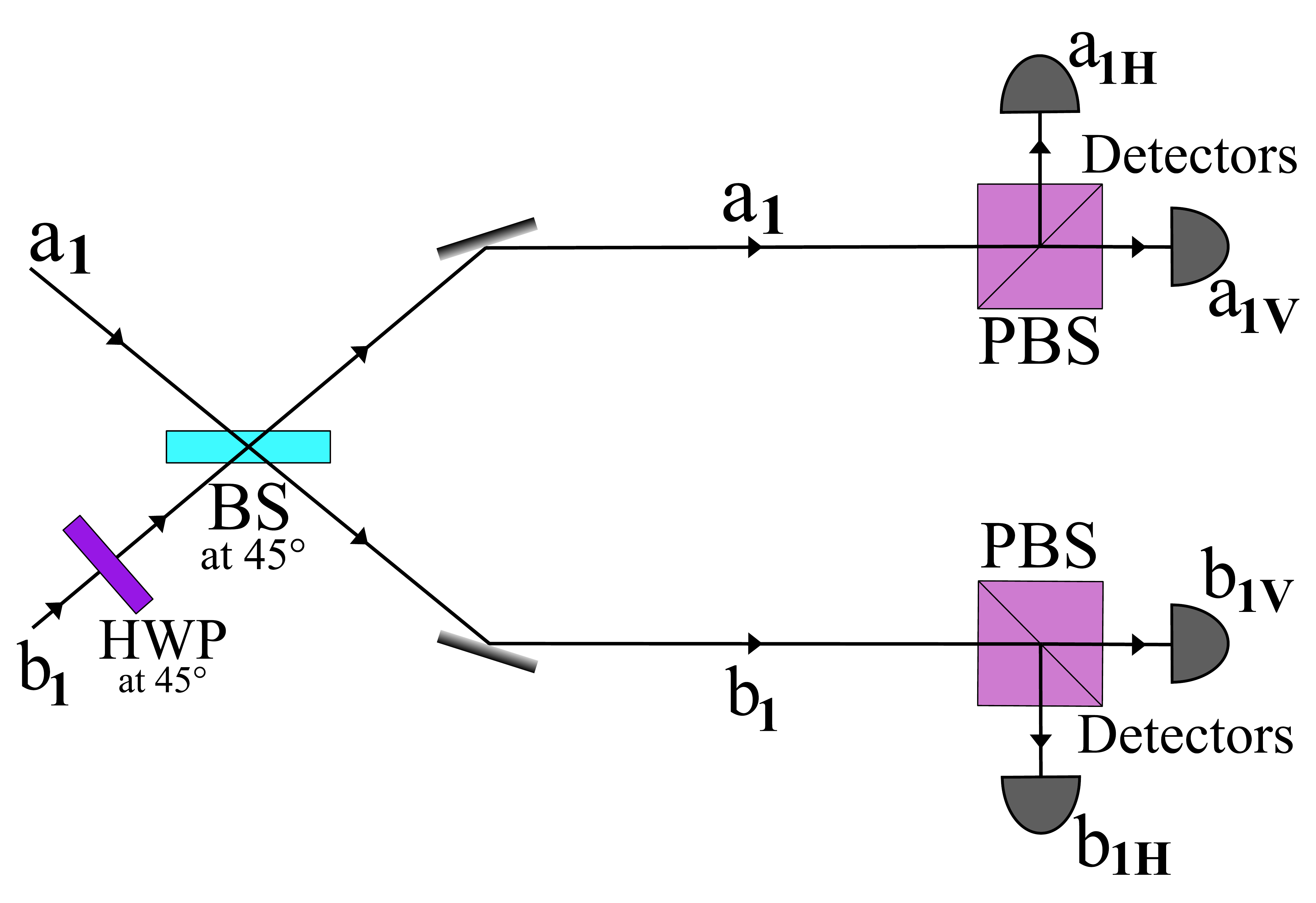}
         \captionlistentry{}
        \label{Fig:1a}
     \end{subfigure}
     \begin{subfigure}[b]{\linewidth}
        \centering
        \includegraphics[width=0.95\linewidth]{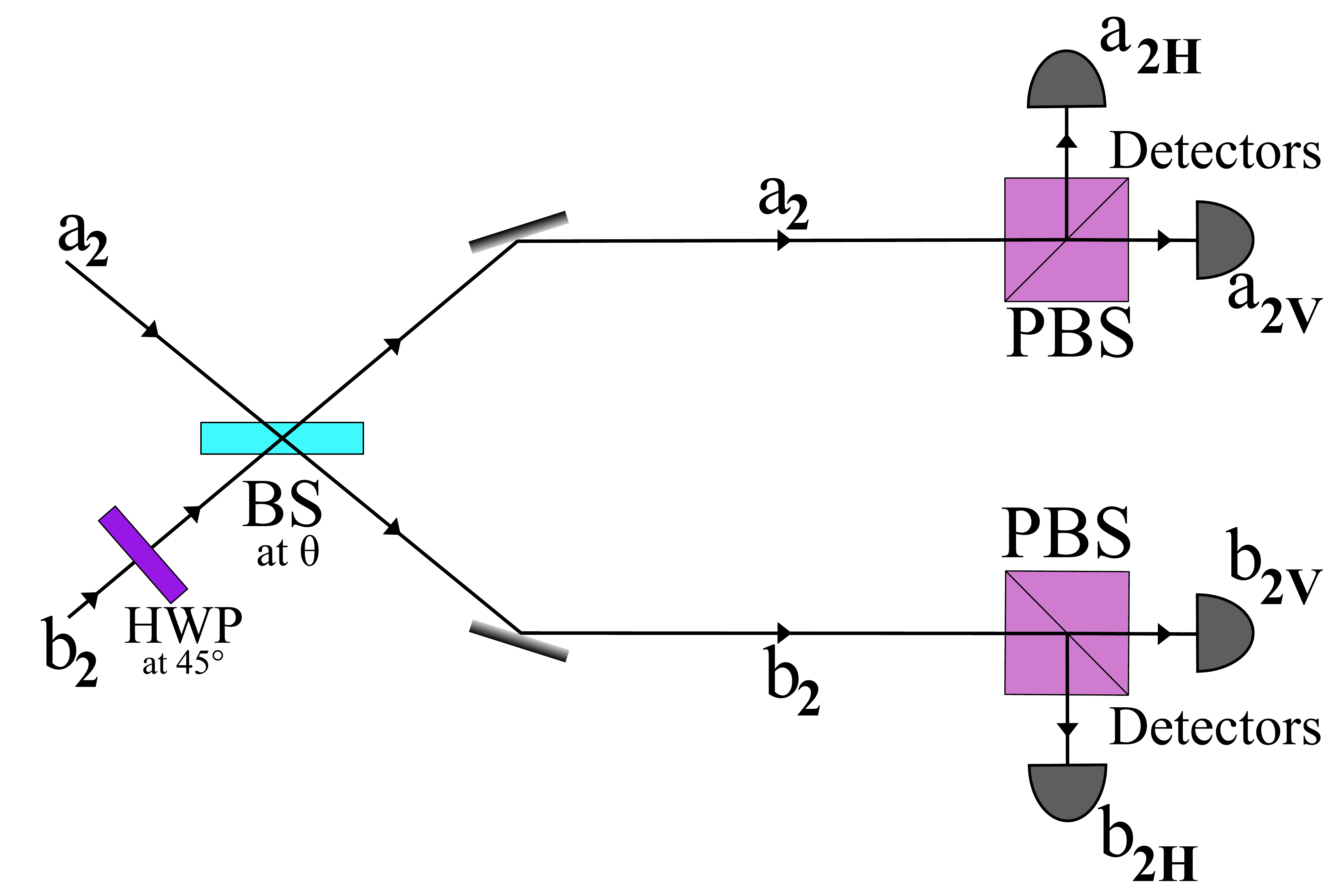}
        \captionlistentry{}
         \label{Fig:1b}
     \end{subfigure}
        \caption{Circuit for the hyper-entangled Bell-like state analyzer where momentum DOF act as system qubits while polarisation DOF act as target qubits.}
        \label{fig:1}
\end{figure}

Thus, we have here the task of distinguishing -- in a linear optical setup -- the four non-maximally entangled states of two photons, given in Eqs. (\ref{5}) -- (\ref{8}). We pass them through the circuit in Fig.\ref{fig:1}. The circuit closely resembles the one used in ref.  \cite{PhysRevA.68.042313}. Two half-wave plates (HWPs) mounted at an incidence angle of $45^{\circ}$ are placed in paths $b_1$ and $b_2$. This scheme differs from the one in \cite{PhysRevA.68.042313} as, instead of using balanced beam splitters in paths of both $(a_1,b_1)$ and $(a_2,b_2)$, here we use a balanced beam splitter in the path $(a_1, b_1)$ while the beam splitter used in the path of $(a_2,b_2)$ is unbalanced, having transmission coefficient $\cos^2(\theta)$. This is in line with the approach in Ref. \cite{PhysRevA.106.023706} for Bell-like state discrimination by introducing an unbalanced beam splitter to tackle the asymmetry of Bell-like states.  This imbalance leads to cancellation of certain unique detection signatures from each of the four Bell-like states, rendering them distinguishable. Then a polarising beam splitter is used, and single-photon detectors are placed at both of its ends.

 In Fig. (\ref{fig:1}), the setup before the beam splitter (BS) containing HWP acts as a CNOT gate where momentum DOFs act as control qubits and polarisation DOFs act as target qubits. The action of this operation can be summarised mathematically through the following transformations:

 \begin{align}
     &\hat{a}^{\dag}_i\hat{h}^{\dag}_i\ket{0}\longrightarrow\hat{a}^{\dag}_i\hat{h}^{\dag}_i\ket{0},\qquad\hat{a}^{\dag}_i\hat{v}^{\dag}_i\ket{0}\longrightarrow\hat{a}^{\dag}_i\hat{v}^{\dag}_i\ket{0},\nonumber\\
     &\hat{b}^{\dag}_i\hat{h}^{\dag}_i\ket{0}\longrightarrow\hat{b}^{\dag}_i\hat{v}^{\dag}_i\ket{0},\qquad~\hat{b}^{\dag}_i\hat{v}^{\dag}_i\ket{0}\longrightarrow\hat{b}^{\dag}_i\hat{h}^{\dag}_i\ket{0},
 \end{align}

with $i=1,2$. Here, the action of HWP mounted at an angle $\theta$ is given as
\begin{align}
    &\hat{h}^{\dag}_i\ket{0}\longrightarrow (\cos(2\theta)\hat{h}^{\dag}_i+\sin(2\theta)\hat{v}^{\dag}_i)\ket{0},\\
    &\hat{v}^{\dag}_i\ket{0}\longrightarrow (\sin(2\theta)\hat{h}^{\dag}_i-\cos(2\theta)\hat{v}^{\dag}_i)\ket{0}.
\end{align}

The action of the balanced beam splitter is given as:
\begin{align}
    &\hat{a}^{\dag}_1\ket{0}\longrightarrow\frac{(\hat{a}^{\dag}_1+\hat{b}^{\dag}_1)}{\sqrt{2}}\ket{0},\qquad\hat{b}^{\dag}_1\ket{0}\longrightarrow\frac{(\hat{a}^{\dag}_1-\hat{b}^{\dag}_1)}{\sqrt{2}}\ket{0}.
\end{align}
Similarly, the action of an unbalanced beam splitter can be written as:
\begin{align}
    &\hat{a}^{\dag}_2\ket{0}\longrightarrow (\cos(\theta)\hat{a}^{\dag}_2+\sin(\theta)\hat{b}^{\dag}_2)\ket{0},\\
    &\hat{b}^{\dag}_2\ket{0}\longrightarrow (\sin(\theta)\hat{a}^{\dag}_2-\cos(\theta)\hat{b}^{\dag}_2)\ket{0}.
\end{align}
 Being equipped with all the necessary transformations, we can now analyze what will happen to the states in Eqs. (\ref{5})-(\ref{8}) when they are passed through the circuit in Fig.\ref{fig:1}. We can easily verify that before entering the PBS for detection, the hyperentangled states are
\begin{align}
    \vert \Theta_1\rangle\longrightarrow\Big(&\frac{1}{2}\hat{a}^{\dag}_1\hat{a}^{\dag}_2\hat{h}^{\dag}_1\hat{v}^{\dag}_2 \sin \left(2\theta \right)+\frac{1}{2}\hat{a}^{\dag}_1\hat{a}^{\dag}_2\hat{h}^{\dag}_2
  \hat{v}^{\dag}_1 \sin \left(2\theta \right)\nonumber\\
   &-\frac{1}{2}\hat{a}^{\dag}_1\hat{b}^{\dag}_2\hat{h}^{\dag}_1\hat{v}^{\dag}_2 \cos \left(2\theta \right)-\frac{1}{2}
  \hat{a}^{\dag}_1\hat{b}^{\dag}_2\hat{h}^{\dag}_2\hat{v}^{\dag}_1\cos \left(2\theta \right)\nonumber\\
   &+\frac{1}{2}\hat{b}^{\dag}_1\hat{b}^{\dag}_2
  \hat{h}^{\dag}_1\hat{v}^{\dag}_2+\frac{1}{2}\hat{b}^{\dag}_1\hat{b}^{\dag}_2\hat{h}^{\dag}_2\hat{v}^{\dag}_1
  \Big)\ket{0},
\end{align}
\begin{align}
    \vert \Theta_2\rangle\longrightarrow\Big(&\frac{1}{2}\hat{a}^{\dag}_1\hat{a}^{\dag}_2\hat{h}^{\dag}_1\hat{v}^{\dag}_2 \cos \left(2\theta \right)+\frac{1}{2}
  \hat{a}^{\dag}_1\hat{a}^{\dag}_2\hat{h}^{\dag}_2\hat{v}^{\dag}_1\cos \left(2\theta \right)\nonumber\\
   &+\frac{1}{2}\hat{a}^{\dag}_1\hat{b}^{\dag}_2\hat{h}^{\dag}_1
  \hat{v}^{\dag}_2 \sin \left(2\theta \right)+\frac{1}{2}\hat{a}^{\dag}_1\hat{b}^{\dag}_2\hat{h}^{\dag}_2\hat{v}^{\dag}_1 \sin
   \left(2\theta \right)\nonumber\\
   &+\frac{1}{2}
  \hat{a}^{\dag}_2\hat{b}^{\dag}_1\hat{h}^{\dag}_1\hat{v}^{\dag}_2 +\frac{1}{2}\hat{a}^{\dag}_2\hat{b}^{\dag}_1
  \hat{h}^{\dag}_2\hat{v}^{\dag}_1\Big)\ket{0},
\end{align}
\begin{align}
    \vert \Theta_3\rangle\longrightarrow\Big(&\frac{1}{2} \hat{a}^{\dag}_1 \hat{a}^{\dag}_2 \hat{h}^{\dag}_1 \hat{h}^{\dag}_2+\frac{1}{2}
   \hat{a}^{\dag}_1 \hat{a}^{\dag}_2 \hat{v}^{\dag}_1 \hat{v}^{\dag}_2 \nonumber\\
   &-\frac{1}{2} \hat{a}^{\dag}_2 \hat{b}^{\dag}_1
   \hat{h}^{\dag}_1 \hat{h}^{\dag}_2 \cos \left(2\theta
   \right)-\frac{1}{2} \hat{a}^{\dag}_2 \hat{b}^{\dag}_1 \hat{v}^{\dag}_1 \hat{v}^{\dag}_2 \cos \left(2\theta \right)\nonumber\\
   &-\frac{1}{2}\hat{b}^{\dag}_1
   \hat{b}^{\dag}_2 \hat{h}^{\dag}_1 \hat{h}^{\dag}_2 \sin \left(2\theta \right) -\frac{1}{2}\hat{b}^{\dag}_1 \hat{b}^{\dag}_2 \hat{v}^{\dag}_1 \hat{v}^{\dag}_2
   \sin \left(2\theta \right) \Big)\ket{0},
\end{align}
\begin{align}
    \vert \Theta_4\rangle\longrightarrow\Big(&-\frac{1}{2}\hat{a}^{\dag}_1\hat{b}^{\dag}_2\hat{h}^{\dag}_1\hat{h}^{\dag}_2 -\frac{1}{2}
  \hat{a}^{\dag}_1\hat{b}^{\dag}_2\hat{v}^{\dag}_1\hat{v}^{\dag}_2 \nonumber\\
   &+\frac{1}{2}\hat{a}^{\dag}_2\hat{b}^{\dag}_1\hat{h}^{\dag}_1
  \hat{h}^{\dag}_2 \sin \left(2\theta \right)+\frac{1}{2}\hat{a}^{\dag}_2\hat{b}^{\dag}_1\hat{v}^{\dag}_1\hat{v}^{\dag}_2 \sin
   \left(2\theta \right)\nonumber\\
   &-\frac{1}{2}
  \hat{b}^{\dag}_1\hat{b}^{\dag}_2\hat{h}^{\dag}_1\hat{h}^{\dag}_2\cos \left(2\theta \right)-\frac{1}{2}\hat{b}^{\dag}_1\hat{b}^{\dag}_2
  \hat{v}^{\dag}_1\hat{v}^{\dag}_2 \cos \left(2\theta
   _1\right)\Big)\ket{0}.
\end{align}
The individual photons in these states then pass through a polarisation beam splitter (PBS) and are detected by single-photon detectors. The PBS transmits the vertically polarised photon while reflecting the horizontally polarised photon. We here make two observations: 
\begin{itemize}
    \item For states $\vert \Theta_1\rangle$ and $\vert \Theta_2\rangle$, both photons are always detected in detectors oriented in directions perpendicular to each other. Furthermore, in this case, we have the following observation:
    \begin{enumerate}
        \item The terms $\{\hat{b}^{\dag}_1\hat{b}^{\dag}_2\hat{v}^{\dag}_1\hat{h}^{\dag}_2,\hat{b}^{\dag}_1\hat{b}^{\dag}_2
  \hat{h}^{\dag}_1\hat{v}^{\dag}_2\}$ are uniquely present in $\vert \Theta_1\rangle$. Thus clicking of detectors $\textbf{b}_{\textbf{1V}}$ and $\textbf{b}_{\textbf{2H}}$ or $\textbf{b}_{\textbf{1H}}$ and $\textbf{b}_{\textbf{2V}}$ indicates unambiguous detection of this state.
  \item The terms $\{\hat{b}^{\dag}_1\hat{a}^{\dag}_2\hat{v}^{\dag}_1\hat{h}^{\dag}_2,\hat{b}^{\dag}_1\hat{a}^{\dag}_2
  \hat{h}^{\dag}_1\hat{v}^{\dag}_2\}$ are uniquely present in $\vert \Theta_2\rangle$. Thus clicking of detectors $\textbf{b}_{\textbf{1V}}$ and $\textbf{a}_{\textbf{2H}}$ or $\textbf{b}_{\textbf{1H}}$ and $\textbf{a}_{\textbf{2V}}$ indicates unambiguous detection of this state.
    \end{enumerate}
    \item For states $\vert \Theta_3\rangle$ and $\vert \Theta_4\rangle$, both photons are always detected in detectors oriented in directions parallel to each other. Furthermore, we have, in this case, the following observation:
    \begin{enumerate}
        \item Terms $\{\hat{a}^{\dag}_1\hat{a}^{\dag}_2\hat{h}^{\dag}_1\hat{h}^{\dag}_2,\hat{a}^{\dag}_1\hat{a}^{\dag}_2
  \hat{v}^{\dag}_1\hat{v}^{\dag}_2\}$ are uniquely present in $\vert \Theta_3\rangle$. Thus clicking of detectors $\textbf{a}_{\textbf{1H}}$ and $\textbf{a}_{\textbf{2H}}$ or $\textbf{a}_{\textbf{1V}}$ and $\textbf{a}_{\textbf{2V}}$ indicates unambiguous detection of this state.
  \item Terms $\{\hat{a}^{\dag}_1\hat{b}^{\dag}_2\hat{h}^{\dag}_1\hat{h}^{\dag}_2,\hat{a}^{\dag}_1\hat{b}^{\dag}_2
  \hat{v}^{\dag}_1\hat{v}^{\dag}_2\}$ are uniquely present in $\vert \Theta_4\rangle$. Thus clicking of detectors $\textbf{a}_{\textbf{1H}}$ and $\textbf{b}_{\textbf{2H}}$ or $\textbf{a}_{\textbf{1V}}$ and $\textbf{b}_{\textbf{2V}}$ indicates unambiguous detection of this state.
    \end{enumerate}
\end{itemize}
There will be detection events where some ambiguity will be present about which state is present. Thus, the success probability is necessarily not $100\%$. If we consider that all four states are provided to us with equal probability, then the success probability is:
\begin{align}
   P_{succ}&=\frac{1}{4}\Big(\frac{1}{4}+\frac{1}{4}\Big)+\frac{1}{4}\Big(\frac{1}{4}+\frac{1}{4}\Big)+\frac{1}{4}\Big(\frac{1}{4}+\frac{1}{4}\Big)+\frac{1}{4}\Big(\frac{1}{4}+\frac{1}{4}\Big),\nonumber\\
    &=\frac{1}{2}.
\end{align}
Thus, the success probability of the aforesaid scheme of unambiguous state discrimination is $50\%$. Comparing it with \cite{PhysRevA.106.023706} we note that not only is the success probability boosted from $25\%$, but also all four Bell-like states can be distinguished (though with $50\%$ success probability), as in that case, only two of them at a time ($\vert \psi_3\rangle$ and $\vert \psi_4\rangle)$ or $(\vert \psi_1\rangle$ and $\vert \psi_2\rangle$) can be distinguished. Thus, hyperentanglement acts as a resource in order to discriminate among them just like it happens in Bell state analysis(BSA), although the success probability doesn't reach $100\%$ as it does for Bell states. It is, of course, important to mention here that the {\it optimal} success probability in the case of unambiguous discrimination of the four Bell-like states (supplied with equal a priori probabilities) of two photons in their momenta degrees of freedom while the two photons are also maximally entangled in their polarization degrees of freedom -- as given in Eqs. (\ref{5}) - (\ref{8}) --   may turn out to be more than $50\%$.
\subsection{Ancillary entanglement in momentum degrees of freedom}\label{sec3}
\begin{figure}
     \centering
     \begin{subfigure}[b]{\linewidth}
         \centering
         \includegraphics[width=0.95\linewidth]{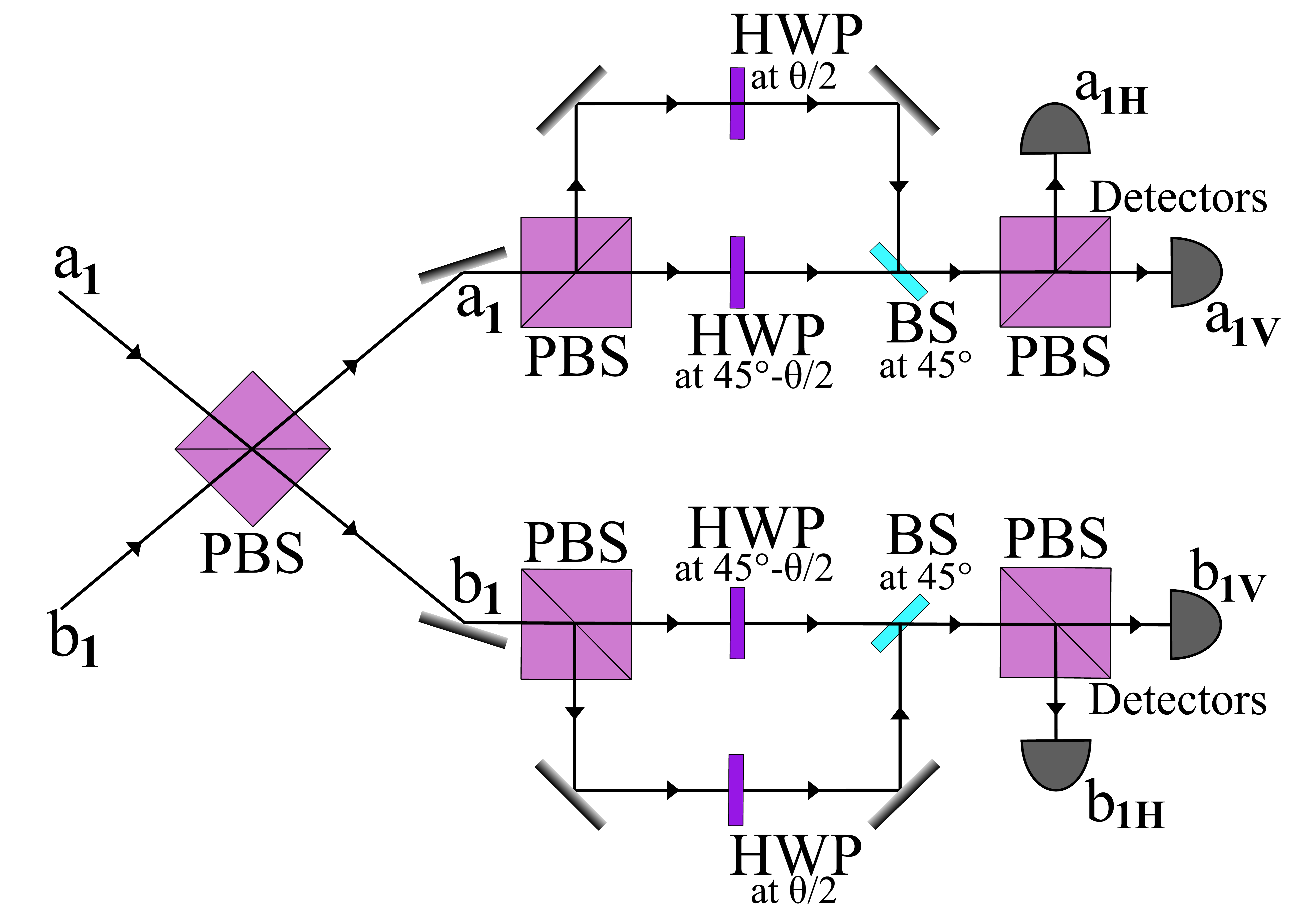}
         \captionlistentry{}
        \label{Fig:2a}
     \end{subfigure}
     \begin{subfigure}[b]{\linewidth}
        \centering
        \includegraphics[width=0.95\linewidth]{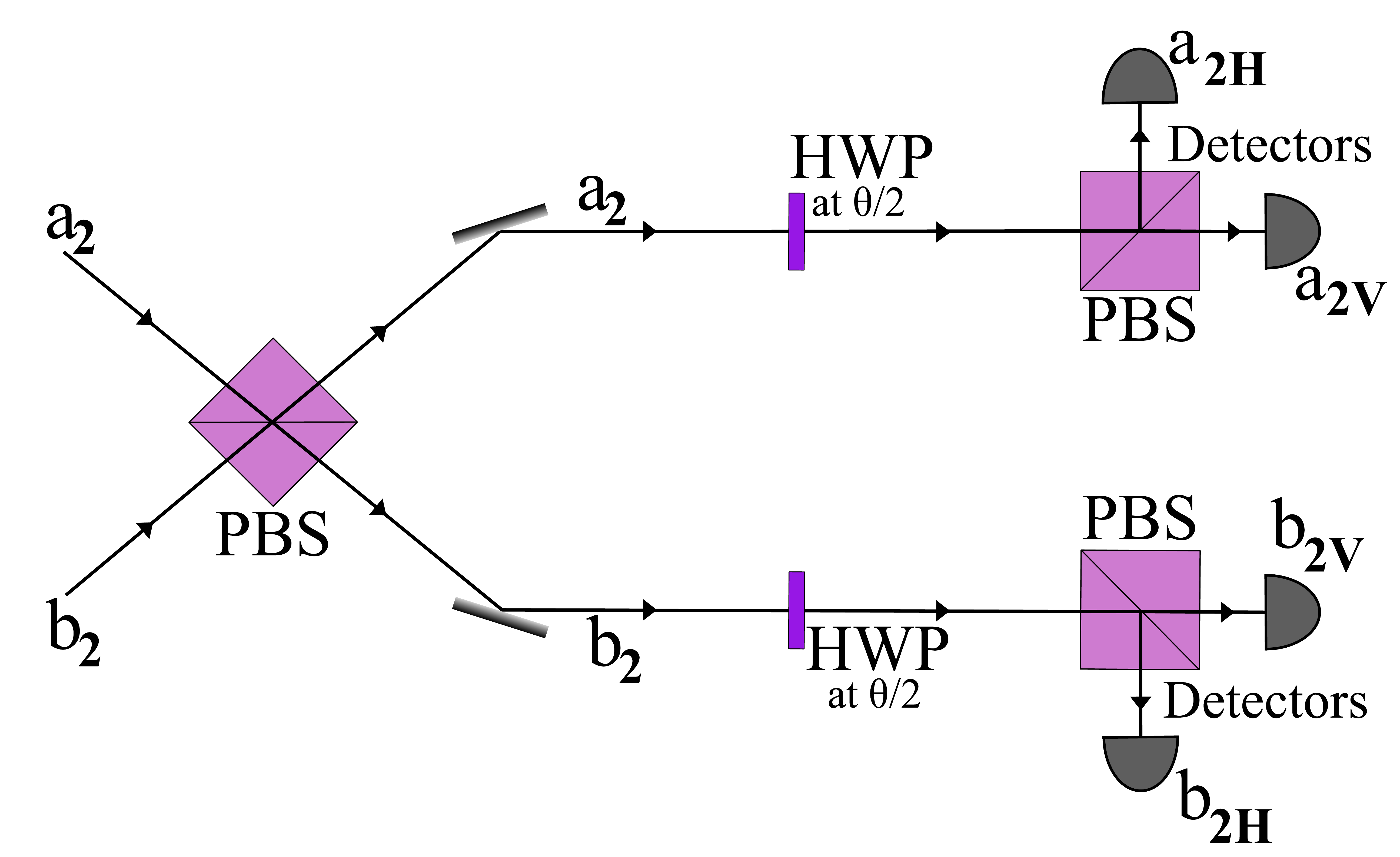}
        \captionlistentry{}
         \label{Fig:2b}
     \end{subfigure}
        \caption{Circuit for the hyper-entangled Bell-like state analyzer where polarisation DOF act as control qubits while momentum DOF act as target qubits. }
        \label{fig:2}
\end{figure}

We can also consider polarisation degrees of freedom as system (i.e control) qubits while momentum degrees of freedom serve as a source of ancillary entanglement. Similar to the previous section, the joint state of four Bell-like states of two-photons in polarisation basis and ancillary qubits in the momentum basis  can be written as
\begin{align}
            &\vert \Xi_1\rangle=(\sin(\theta)\hat{h}^{\dag}_1\hat{h}^{\dag}_2+\cos(\theta)\hat{v}^{\dag}_1\hat{v}^{\dag}_2)\otimes\frac{(\hat{a}^{\dag}_1\hat{b}^{\dag}_2+\hat{b}^{\dag}_1\hat{a}^{\dag}_2)}{\sqrt{2}}\ket{0},\label{34}\\  
            &\vert \Xi_2\rangle=(\cos(\theta)\hat{h}^{\dag}_1\hat{h}^{\dag}_2-\sin(\theta)\hat{v}^{\dag}_1\hat{v}^{\dag}_2)\otimes\frac{(\hat{a}^{\dag}_1\hat{b}^{\dag}_2+\hat{b}^{\dag}_1\hat{a}^{\dag}_2)}{\sqrt{2}}\ket{0},\\
            &\vert \Xi_3\rangle=(\sin(\theta)\hat{h}^{\dag}_1\hat{v}^{\dag}_2+\cos(\theta)\hat{v}^{\dag}_1\hat{h}^{\dag}_2)\otimes\frac{(\hat{a}^{\dag}_1\hat{b}^{\dag}_2+\hat{b}^{\dag}_1\hat{a}^{\dag}_2)}{\sqrt{2}}\ket{0},\\          
            &\vert \Xi_4\rangle=(\cos(\theta)\hat{h}^{\dag}_1\hat{v}^{\dag}_2-\sin(\theta)\hat{v}^{\dag}_1\hat{h}^{\dag}_2)\otimes\frac{(\hat{a}^{\dag}_1\hat{b}^{\dag}_2+\hat{b}^{\dag}_1\hat{a}^{\dag}_2)}{\sqrt{2}}\ket{0}.\label{38}
\end{align}
The circuit used for state discrimination is shown in Fig.\ref{fig:2}. Here, first, photons pass through PBS, which acts as a CNOT operation with polarisation DOF as the control qubit and momentum DOF as the target qubit. The transformation relations are given as:

\begin{align}
     &\hat{h}^{\dag}_i\hat{a}^{\dag}_i\ket{0}\longrightarrow\hat{h}^{\dag}_i\hat{a}^{\dag}_i\ket{0},\qquad\hat{h}^{\dag}_i\hat{b}^{\dag}_i\ket{0}\longrightarrow \hat{h}^{\dag}_i \hat{b}^{\dag}_i\ket{0},\\
      &\hat{v}^{\dag}_i\hat{a}^{\dag}_i\ket{0}\longrightarrow\hat{v}^{\dag}_i\hat{b}^{\dag}_i\ket{0},\qquad~\hat{v}^{\dag}_i\hat{b}^{\dag}_i\ket{0}\longrightarrow\hat{v}^{\dag}_i\hat{a}^{\dag}_i\ket{0},
 \end{align}
 
where again $i=1,2$. After that, another PBS is used in the circuit involving modes $\hat{a}^{\dag}_1$ and $\hat{b}^{\dag}_1$, which transmits the vertically polarised photon while reflecting the horizontally polarised photon. It is used so that HWP mounted at angle $\frac{\theta}{2}$ acts on mode $\hat{h}^{\dag}_1$ while HWP mounted at angle $45^{\circ}-\frac{\theta}{2}$ acts on mode $\hat{v}^{\dag}_1$. As mentioned earlier, this kind of asymmetry is necessary for certain cancellations which render Bell-like states unambiguously distinguishable. In the circuit involving modes $\hat{a}^{\dag}_2$ and $\hat{b}^{\dag}_2$ both are passed through HWP mounted at an angle $\frac{\theta}{2}$. The transformation rules for a modes $\hat{h}^{\dag}_2$ and $\hat{v}^{\dag}_2$are
\begin{align}
    &\hat{h}^{\dag}_2\ket{0}\longrightarrow (\cos(\theta)\hat{h}^{\dag}_2+\sin(\theta)\hat{v}^{\dag}_2)\ket{0},\\
    &\hat{v}^{\dag}_2\ket{0}\longrightarrow (\sin(\theta)\hat{h}^{\dag}_2-\cos(\theta)\hat{v}^{\dag}_2)\ket{0}.
\end{align}
While, modes $\hat{h}^{\dag}_1$ and $\hat{v}^{\dag}_1$ transform as
\begin{align}
    &\hat{h}^{\dag}_1\ket{0}\longrightarrow (\cos(\theta)\hat{h}^{\dag}_1+\sin(\theta)\hat{v}^{\dag}_1)\ket{0},\\
    &\hat{v}^{\dag}_1\ket{0}\longrightarrow (\cos(\theta)\hat{h}^{\dag}_1-\sin(\theta)\hat{v}^{\dag}_1)\ket{0}.
\end{align}
A beam splitter is then used to merge the split modes $\hat{h}^{\dag}_1$ and $\hat{v}^{\dag}_1$ after HWPs act on them. The output is recorded from only one arm of the beam splitter, resulting in a $50\%$ loss. This can be compensated by putting a detector in the other arm of the beam splitter. It is worth mentioning that as only a single photon at a time reaches one arm of the beam splitter, collecting output from either arm of the beam splitter is equivalent from the state discrimination point of view. Collecting output from both arms together will not change the success probability. Thus, for simplicity, we won't consider that case here as our purpose is only to calculate the success probability. After passing through the circuit, the states from Eq. (\ref{34})--- (\ref{38}) just before entering the PBS near detectors are given as:
\begin{align}
    \vert \Xi_1\rangle\longrightarrow\Big(&\frac{\hat{a}^{\dag}_1 \hat{b}^{\dag}_2 \hat{h}^{\dag}_1 \hat{h}^{\dag}_2 \sin \left(\theta\right)}{\sqrt{2}}+\frac{\hat{a}^{\dag}_1 \hat{b}^{\dag}_2 \hat{h}^{\dag}_1 \hat{h}^{\dag}_2 \sin \left(\theta\right) \cos \left(2
   \theta\right)}{\sqrt{2}}\nonumber\\
   &-\frac{\hat{a}^{\dag}_1 \hat{b}^{\dag}_2 \hat{h}^{\dag}_1 \hat{v}^{\dag}_2 \cos \left(\theta\right)}{2 \sqrt{2}}-\frac{\hat{a}^{\dag}_1 \hat{b}^{\dag}_2 \hat{h}^{\dag}_1 \hat{v}^{\dag}_2 \cos
   \left(3 \theta\right)}{2 \sqrt{2}}\nonumber\\
   &+\frac{\hat{a}^{\dag}_1 \hat{b}^{\dag}_2 \hat{v}^{\dag}_1 \hat{v}^{\dag}_2 \sin \left(\theta\right)}{\sqrt{2}}+\frac{\hat{a}^{\dag}_2 \hat{b}^{\dag}_1 \hat{h}^{\dag}_1 \hat{h}^{\dag}_2
   \sin \left(\theta\right)}{\sqrt{2}}\nonumber\\
   &+\frac{\hat{a}^{\dag}_2 \hat{b}^{\dag}_1 \hat{h}^{\dag}_1 \hat{h}^{\dag}_2 \sin \left(\theta\right) \cos \left(2 \theta\right)}{\sqrt{2}}-\frac{\hat{a}^{\dag}_2
   \hat{b}^{\dag}_1 \hat{h}^{\dag}_1 \hat{v}^{\dag}_2 \cos \left(\theta\right)}{2 \sqrt{2}}\nonumber\\
   &-\frac{\hat{a}^{\dag}_2 \hat{b}^{\dag}_1 \hat{h}^{\dag}_1 \hat{v}^{\dag}_2 \cos \left(3 \theta\right)}{2
   \sqrt{2}}+\frac{\hat{a}^{\dag}_2 \hat{b}^{\dag}_1 \hat{v}^{\dag}_1 \hat{v}^{\dag}_2 \sin \left(\theta\right)}{\sqrt{2}}
  \Big)\ket{0},
\end{align}
\begin{align}
    \vert \Xi_2\rangle\longrightarrow\Big(&\frac{\hat{a}^{\dag}_1 \hat{b}^{\dag}_2 \hat{h}^{\dag}_1 \hat{h}^{\dag}_2 \cos \left(\theta\right)}{2 \sqrt{2}}+\frac{\hat{a}^{\dag}_1 \hat{b}^{\dag}_2 \hat{h}^{\dag}_1 \hat{h}^{\dag}_2 \cos \left(3 \theta\right)}{2
   \sqrt{2}}\nonumber\\
   &+\frac{\hat{a}^{\dag}_1 \hat{b}^{\dag}_2 \hat{h}^{\dag}_1 \hat{v}^{\dag}_2 \sin \left(\theta\right)}{\sqrt{2}}+\frac{\hat{a}^{\dag}_1 \hat{b}^{\dag}_2 \hat{h}^{\dag}_1 \hat{v}^{\dag}_2 \sin \left(\theta\right)
   \cos \left(2 \theta\right)}{\sqrt{2}}\nonumber\\
   &+\frac{\hat{a}^{\dag}_1 \hat{b}^{\dag}_2 \hat{h}^{\dag}_2 \hat{v}^{\dag}_1 \sin \left(\theta\right)}{\sqrt{2}}+\frac{\hat{a}^{\dag}_2 \hat{b}^{\dag}_1 \hat{h}^{\dag}_1 \hat{h}^{\dag}_2
   \cos \left(\theta\right)}{2 \sqrt{2}}\nonumber\\
   &+\frac{\hat{a}^{\dag}_2 \hat{b}^{\dag}_1 \hat{h}^{\dag}_1 \hat{h}^{\dag}_2 \cos \left(3 \theta\right)}{2 \sqrt{2}}+\frac{\hat{a}^{\dag}_2 \hat{b}^{\dag}_1 \hat{h}^{\dag}_1
   \hat{v}^{\dag}_2 \sin \left(\theta\right)}{\sqrt{2}}\nonumber\\
   &+\frac{\hat{a}^{\dag}_2 \hat{b}^{\dag}_1 \hat{h}^{\dag}_1 \hat{v}^{\dag}_2 \sin \left(\theta\right) \cos \left(2 \theta
   _1\right)}{\sqrt{2}}+\frac{\hat{a}^{\dag}_2 \hat{b}^{\dag}_1 \hat{h}^{\dag}_2 \hat{v}^{\dag}_1 \sin \left(\theta\right)}{\sqrt{2}}\Big)\ket{0},
\end{align}
\begin{align}
    \vert \Xi_3\rangle\longrightarrow\Big(&\frac{\hat{a}^{\dag}_1 \hat{a}^{\dag}_2 \hat{h}^{\dag}_1 \hat{h}^{\dag}_2 \cos \left(\theta\right)}{\sqrt{2}}+\frac{\hat{a}^{\dag}_1 \hat{a}^{\dag}_2 \hat{h}^{\dag}_2 \hat{v}^{\dag}_1 \sin \left(\theta\right)}{2
   \sqrt{2}}\nonumber\\
   &-\frac{\hat{a}^{\dag}_1 \hat{a}^{\dag}_2 \hat{h}^{\dag}_2 \hat{v}^{\dag}_1 \sin \left(3 \theta\right)}{2 \sqrt{2}}-\frac{\hat{a}^{\dag}_1 \hat{a}^{\dag}_2 \hat{v}^{\dag}_1 \hat{v}^{\dag}_2 \cos \left(\theta
   _1\right)}{2 \sqrt{2}}\nonumber\\
   &+\frac{\hat{a}^{\dag}_1 \hat{a}^{\dag}_2 \hat{v}^{\dag}_1 \hat{v}^{\dag}_2 \cos \left(3 \theta\right)}{2 \sqrt{2}}+\frac{\hat{b}^{\dag}_1 \hat{b}^{\dag}_2 \hat{h}^{\dag}_1 \hat{h}^{\dag}_2 \cos
   \left(\theta\right)}{\sqrt{2}}\nonumber\\
   &+\frac{\hat{b}^{\dag}_1 \hat{b}^{\dag}_2 \hat{h}^{\dag}_2 \hat{v}^{\dag}_1 \sin \left(\theta\right)}{2 \sqrt{2}}-\frac{\hat{b}^{\dag}_1 \hat{b}^{\dag}_2 \hat{h}^{\dag}_2 \hat{v}^{\dag}_1 \sin
   \left(3 \theta\right)}{2 \sqrt{2}}\nonumber\\
   &-\frac{\hat{b}^{\dag}_1 \hat{b}^{\dag}_2 \hat{v}^{\dag}_1 \hat{v}^{\dag}_2 \cos \left(\theta\right)}{2 \sqrt{2}}+\frac{\hat{b}^{\dag}_1 \hat{b}^{\dag}_2 \hat{v}^{\dag}_1 \hat{v}^{\dag}_2
   \cos \left(3 \theta\right)}{2 \sqrt{2}}\Big)\ket{0},
\end{align}
\begin{align}
    \vert \Xi_4\rangle\longrightarrow\Big(&-\frac{\hat{a}^{\dag}_1 \hat{a}^{\dag}_2 \hat{h}^{\dag}_1 \hat{v}^{\dag}_2 \cos \left(\theta\right)}{\sqrt{2}}+\frac{\hat{a}^{\dag}_1 \hat{a}^{\dag}_2 \hat{h}^{\dag}_2 \hat{v}^{\dag}_1 \cos \left(\theta\right)}{2
   \sqrt{2}}\nonumber\\
   &-\frac{\hat{a}^{\dag}_1 \hat{a}^{\dag}_2 \hat{h}^{\dag}_2 \hat{v}^{\dag}_1 \cos \left(3 \theta\right)}{2 \sqrt{2}}+\frac{\hat{a}^{\dag}_1 \hat{a}^{\dag}_2 \hat{v}^{\dag}_1 \hat{v}^{\dag}_2 \sin \left(\theta
   _1\right)}{2 \sqrt{2}}\nonumber\\
     &-\frac{\hat{a}^{\dag}_1 \hat{a}^{\dag}_2 \hat{v}^{\dag}_1 \hat{v}^{\dag}_2 \sin \left(3 \theta\right)}{2 \sqrt{2}}-\frac{\hat{b}^{\dag}_1 \hat{b}^{\dag}_2 \hat{h}^{\dag}_1 \hat{v}^{\dag}_2 \cos
   \left(\theta\right)}{\sqrt{2}}\nonumber\\
   &+\frac{\hat{b}^{\dag}_1 \hat{b}^{\dag}_2 \hat{h}^{\dag}_2 \hat{v}^{\dag}_1 \cos \left(\theta\right)}{2 \sqrt{2}}-\frac{\hat{b}^{\dag}_1 \hat{b}^{\dag}_2 \hat{h}^{\dag}_2 \hat{v}^{\dag}_1 \cos
   \left(3 \theta\right)}{2 \sqrt{2}}\nonumber\\
   &+\frac{\hat{b}^{\dag}_1 \hat{b}^{\dag}_2 \hat{v}^{\dag}_1 \hat{v}^{\dag}_2 \sin \left(\theta\right)}{2 \sqrt{2}}-\frac{\hat{b}^{\dag}_1 \hat{b}^{\dag}_2 \hat{v}^{\dag}_1 \hat{v}^{\dag}_2
   \sin \left(3 \theta\right)}{2 \sqrt{2}}\Big)\ket{0}.
\end{align}
We make the following observations
\begin{itemize}
    \item For states $\vert \Xi_1\rangle$ and $\vert \Xi_2\rangle$ both the photons are always detected either in modes $\{\hat{a}^{\dag}_1,\hat{b}^{\dag}_2\}$ or $\{\hat{b}^{\dag}_1,\hat{a}^{\dag}_2\}$. Furthermore between them
    \begin{enumerate}
        \item Terms $\{\hat{b}^{\dag}_1\hat{a}^{\dag}_2\hat{v}^{\dag}_2\hat{v}^{\dag}_1,\hat{a}^{\dag}_1\hat{b}^{\dag}_2
  \hat{v}^{\dag}_1\hat{v}^{\dag}_2\}$ are uniquely present in $\vert \Xi_1\rangle$. Thus clicking of detectors $\textbf{b}_{\textbf{1V}}$ and $\textbf{a}_{\textbf{2V}}$ or $\textbf{a}_{\textbf{1V}}$ and $\textbf{b}_{\textbf{2V}}$ indicates unambiguous detection of this state.
  \item Terms $\{\hat{b}^{\dag}_1\hat{a}^{\dag}_2\hat{h}^{\dag}_2\hat{v}^{\dag}_1,\hat{a}^{\dag}_1\hat{b}^{\dag}_2
  \hat{h}^{\dag}_1\hat{v}^{\dag}_2\}$ are uniquely present in $\vert \Xi_2\rangle$. Thus clicking of detectors $\textbf{a}_{\textbf{1V}}$ and $\textbf{b}_{\textbf{2H}}$ or $\textbf{b}_{\textbf{1H}}$ and $\textbf{a}_{\textbf{2V}}$ indicates unambiguous detection of this state.
    \end{enumerate}
    \item For states $\vert \Xi_3\rangle$ and $\vert \Xi_4\rangle$ both the photons are always detected as either in modes $\{\hat{a}^{\dag}_1,\hat{a}^{\dag}_2\}$ or $\{\hat{b}^{\dag}_1,\hat{b}^{\dag}_2\}$.
    \begin{enumerate}
        \item Terms $\{\hat{a}^{\dag}_1\hat{a}^{\dag}_2\hat{h}^{\dag}_2\hat{h}^{\dag}_1,\hat{b}^{\dag}_1\hat{b}^{\dag}_2
  \hat{h}^{\dag}_1\hat{h}^{\dag}_2\}$ are uniquely present in $\vert \Xi_3\rangle$. Thus clicking of detectors $\textbf{a}_{\textbf{1H}}$ and $\textbf{a}_{\textbf{2H}}$ or $\textbf{b}_{\textbf{1V}}$ and $\textbf{b}_{\textbf{2V}}$ indicates unambiguous detection of this state.
  \item Terms $\{\hat{a}^{\dag}_1\hat{a}^{\dag}_2\hat{v}^{\dag}_2\hat{h}^{\dag}_1,\hat{b}^{\dag}_1\hat{b}^{\dag}_2
  \hat{h}^{\dag}_1\hat{v}^{\dag}_2\}$ are uniquely present in $\vert \Xi_4\rangle$. Thus clicking of detectors $\textbf{a}_{\textbf{1H}}$ and $\textbf{a}_{\textbf{2V}}$ or $\textbf{b}_{\textbf{1H}}$ and $\textbf{b}_{\textbf{2V}}$ indicates unambiguous detection of this state.
    \end{enumerate}
\end{itemize}
Again, there will be some detections that will render the discrimination of these states ambiguous. Thus, the success probability again won't be $100\%$. It can be calculated in a similar way to the previous section and comes out to be
\begin{align}
   P_{succ}&=\frac{1}{4}(\frac{\sin^2(\theta)}{2}+\frac{\sin^2(\theta)}{2})+\frac{1}{4}(\frac{\sin^2(\theta)}{2}+\frac{\sin^2(\theta)}{2}),\nonumber\\
    &+\frac{1}{4}(\frac{\cos^2(\theta)}{2}+\frac{\cos^2(\theta)}{2})+\frac{1}{4}(\frac{\cos^2(\theta)}{2}+\frac{\cos^2(\theta)}{2}).\nonumber\\
    &=\frac{1}{2}
\end{align}
Thus, in this case, we see that the success probability of unambiguous state discrimination is again $50\%$.

\subsection{Ancillary entanglement in OAM degrees of freedom}\label{sec4}
\begin{figure}
     \centering
     \begin{subfigure}[b]{\linewidth}
         \centering
         \includegraphics[width=1\linewidth]{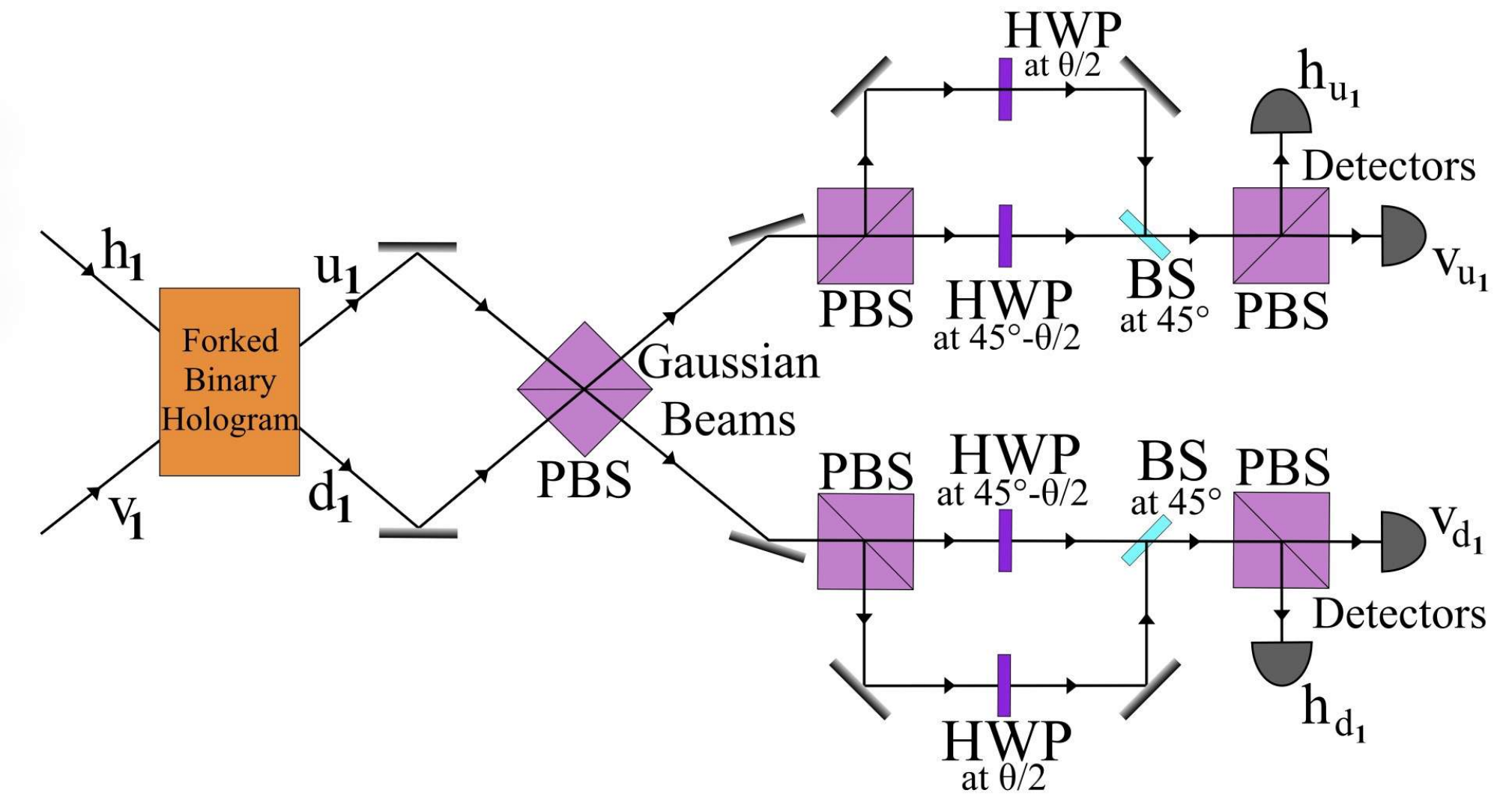}
         \captionlistentry{}
        \label{Fig:3a}
     \end{subfigure}
     \begin{subfigure}[b]{1\linewidth}
        \centering
        \includegraphics[width=1\linewidth]{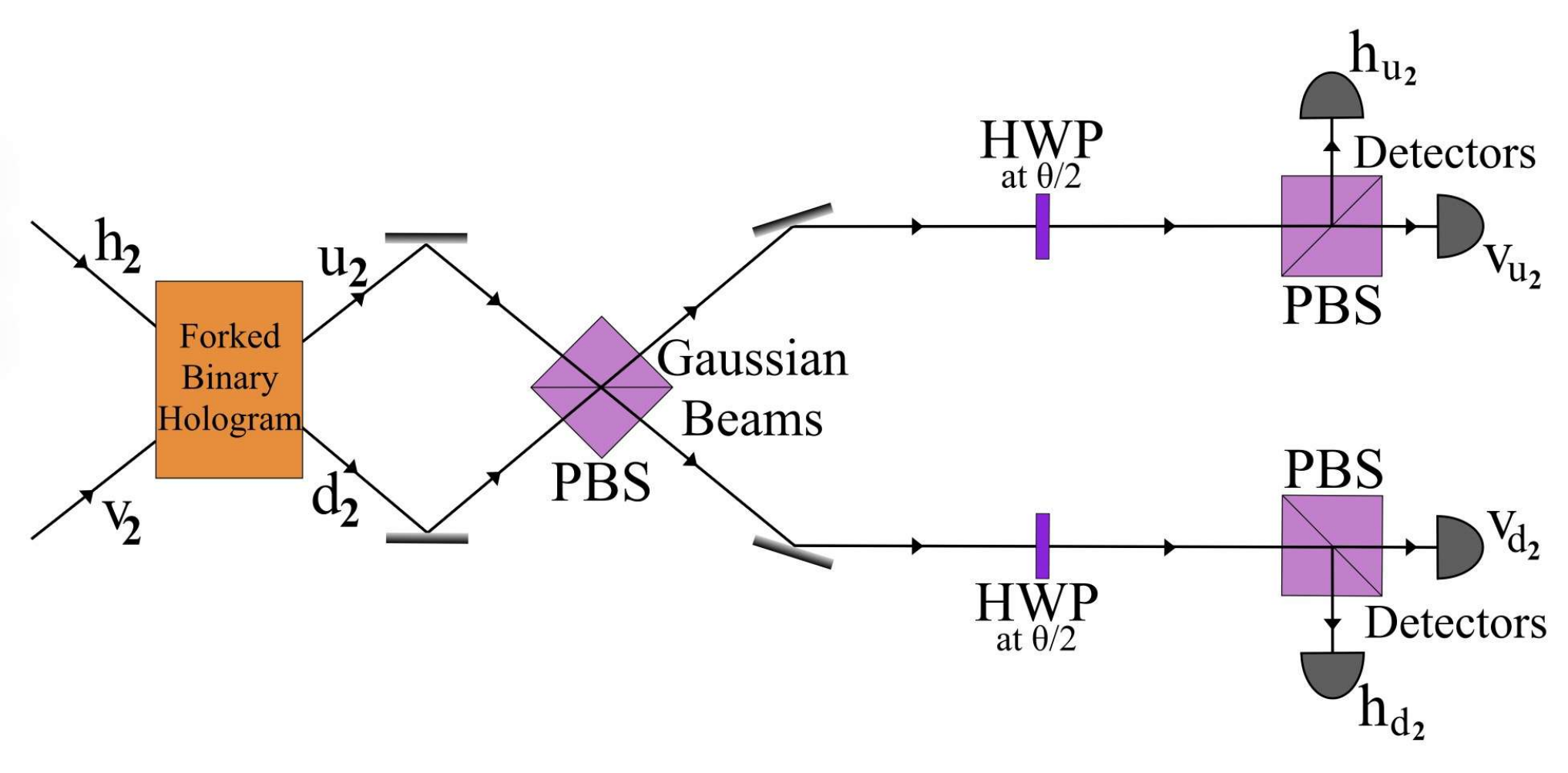}
        \captionlistentry{}
         \label{Fig:3b}
     \end{subfigure}
        \caption{Circuit for the hyper-entangled Bell-like state analyzer where polarisation DOF act as system qubits while OAM DOF act as ancillary qubits.}
        \label{fig:3}
\end{figure}

Till now, we have discussed how momentum DOF act as system (ancilla) qubits while polarisation DOF acts as ancillary (system) qubits to assist in the unambiguous discrimination of Bell-like states. But photons have orbital angular momentum (OAM) DOF also. They can act as maximally entangled ancillary qubits also while polarisation DOF being the system qubits. Such a scheme has been adopted by Kwiat et.al. in \cite{barreiro2008beating} to implement super-dense coding. The circuit presented here is on the same line as that of \cite{barreiro2008beating}. It is shown in Fig. (\ref{fig:3})

The joint initial states of the system and ancillary qubits are given as:
\begin{align}\label{Belllikehypero}
            &\vert \Pi_1\rangle=(\sin(\theta)\hat{h}^{\dag}_1\hat{h}^{\dag}_2+\cos(\theta)\hat{v}^{\dag}_1\hat{v}^{\dag}_2)\otimes\frac{(\hat{p}^{\dag}_1\hat{m}^{\dag}_2+\hat{m}^{\dag}_1\hat{p}^{\dag}_2)}{\sqrt{2}}\ket{0}\\  
            &\vert \Pi_2\rangle=(\cos(\theta)\hat{h}^{\dag}_1\hat{h}^{\dag}_2-\sin(\theta)\hat{v}^{\dag}_1\hat{v}^{\dag}_2)\otimes\frac{(\hat{p}^{\dag}_1\hat{m}^{\dag}_2+\hat{m}^{\dag}_1\hat{p}^{\dag}_2)}{\sqrt{2}}\ket{0}\\
            &\vert \Pi_3\rangle=(\sin(\theta)\hat{h}^{\dag}_1\hat{v}^{\dag}_2+\cos(\theta)\hat{v}^{\dag}_1\hat{h}^{\dag}_2)\otimes\frac{(\hat{p}^{\dag}_1\hat{m}^{\dag}_2+\hat{m}^{\dag}_1\hat{p}^{\dag}_2)}{\sqrt{2}}\ket{0}\\        
            &\vert \Pi_4\rangle=(\cos(\theta)\hat{h}^{\dag}_1\hat{v}^{\dag}_2-\sin(\theta)\hat{v}^{\dag}_1\hat{h}^{\dag}_2)\otimes\frac{(\hat{p}^{\dag}_1\hat{m}^{\dag}_2+\hat{m}^{\dag}_1\hat{p}^{\dag}_2)}{\sqrt{2}}\ket{0}\label{388}  
\end{align}
where $\hat{p}^{\dag}$ represent creation operator for OAM mode $+1$ while  $\hat{m}^{\dag}$ represent creation operator for OAM mode $-1$. When the i-th photon with OAM value $+ 1  (- 1)$ falls onto the Forked Binary Hologram, it will take the upper (down) path designated by the mode $u_i$ ($d_i$) -- for i = 1, 2. Then they pass through an assembly of PBSs and HWPs and are finally detected. To discriminate among Bell-like states, it is necessary that HWPs act differently on both horizontally and vertically polarised first photon. Hence, a PBS is used to split these two and then merge again by only recording output from one arm of a balanced beam splitter. This costs us a $50\%$ loss in the beam, which, as mentioned in the previous section, can be compensated by putting the detector in the other arm also without changing the success probability. This is because of the fact that in this setup also only one photon at a time reaches the port of the beam splitter. The action of the forked binary hologram is given as:
\begin{align}
    &\hat{h}^{\dag}_i\hat{p}^{\dag}_i\longrightarrow\hat{h}_{ u_i}^{\dag}\hat{g}^{\dag}_i,\qquad\hat{h}^{\dag}_i\hat{m}^{\dag}_i\longrightarrow\hat{h}_{ d_i}^{\dag}\hat{g}^{\dag}_i,\\
    &\hat{v}^{\dag}_i\hat{p}^{\dag}_i\longrightarrow\hat{v}_{ u_i}^{\dag}\hat{g}^{\dag}_i,\qquad\hat{v}^{\dag}_i\hat{m}^{\dag}_i\longrightarrow\hat{v}_{ d_i}^{\dag}\hat{g}^{\dag}_i,
\end{align}
where $\hat{g}^{\dag}_i$ represents the creation operator for the mode of the $ i$th photon with $0$ OAM value and $\hat{h}(\hat{v})_{ u_i(d_i)}^{\dag}$ represents the horizontally (vertically) polarised $i$th photon in upper (down) path, where $i=1,2.$ Thus, the final states just before the final detection at PBS become
\begin{align}
     \vert \Pi_1\rangle\longrightarrow\Big(&\frac{\hat{h}_{d_1}^{\dag } \hat{h}_{d_2}^{\dag } \cos \left(\theta\right)}{\sqrt{2}}+\frac{\hat{h}_{u_1}^{\dag } \hat{h}_{u_2}^{\dag } \cos \left(\theta\right)}{\sqrt{2}}\nonumber\\
     &+\frac{\hat{h}_{d_2}^{\dag } \hat{v}_{d_1}^{\dag } \sin
   \left(\theta\right)}{2 \sqrt{2}}-\frac{\hat{h}_{d_2}^{\dag } \hat{v}_{d_1}^{\dag } \sin \left(3 \theta\right)}{2 \sqrt{2}}\nonumber\\
   &+\frac{\hat{h}_{u_2}^{\dag } \hat{v}_{u_1}^{\dag } \sin \left(\theta\right)}{2
   \sqrt{2}}-\frac{\hat{h}_{u_2}^{\dag } \hat{v}_{u_1}^{\dag } \sin \left(3 \theta\right)}{2 \sqrt{2}}\nonumber\\
   &-\frac{\hat{v}_{d_1}^{\dag } \hat{v}_{d_2}^{\dag } \cos \left(\theta\right)}{2 \sqrt{2}}+\frac{\hat{v}_{d_1}^{\dag }
   \hat{v}_{d_2}^{\dag } \cos \left(3 \theta\right)}{2 \sqrt{2}}\nonumber\\
   &-\frac{\hat{v}_{u_1}^{\dag } \hat{v}_{u_2}^{\dag } \cos \left(\theta\right)}{2 \sqrt{2}}+\frac{\hat{v}_{u_1}^{\dag } \hat{v}_{u_2}^{\dag } \cos \left(3
   \theta\right)}{2 \sqrt{2}}\Big)\otimes\hat{g}^{\dag}_1\hat{g}^{\dag}_2\ket{0},\label{39}
\end{align}
\begin{align}
     \vert \Pi_2\rangle\longrightarrow\Big(&-\frac{\hat{h}_{d_1}^{\dag } \hat{v}_{d_2}^{\dag } \cos \left(\theta\right)}{\sqrt{2}}-\frac{\hat{h}_{u_1}^{\dag } \hat{v}_{u_2}^{\dag } \cos \left(\theta\right)}{\sqrt{2}}\nonumber\\
     &+\frac{\hat{h}_{d_2}^{\dag } \hat{v}_{d_1}^{\dag } \cos
   \left(\theta\right)}{2 \sqrt{2}}-\frac{\hat{h}_{d_2}^{\dag } \hat{v}_{d_1}^{\dag } \cos \left(3 \theta\right)}{2 \sqrt{2}}\nonumber\\
   &+\frac{\hat{h}_{u_2}^{\dag } \hat{v}_{u_1}^{\dag } \cos \left(\theta\right)}{2
   \sqrt{2}}-\frac{\hat{h}_{u_2}^{\dag } \hat{v}_{u_1}^{\dag } \cos \left(3 \theta\right)}{2 \sqrt{2}}\nonumber\\
   &+\frac{\hat{v}_{d_1}^{\dag } \hat{v}_{d_2}^{\dag } \sin \left(\theta\right)}{2 \sqrt{2}}-\frac{\hat{v}_{d_1}^{\dag }
   \hat{v}_{d_2}^{\dag } \sin \left(3 \theta\right)}{2 \sqrt{2}}\nonumber\\
   &+\frac{\hat{v}_{u_1}^{\dag } \hat{v}_{u_2}^{\dag } \sin \left(\theta\right)}{2 \sqrt{2}}-\frac{\hat{v}_{u_1}^{\dag } \hat{v}_{u_2}^{\dag } \sin \left(3
   \theta\right)}{2 \sqrt{2}}\Big)\otimes\hat{g}^{\dag}_1\hat{g}^{\dag}_2\ket{0},\label{40}
\end{align}
\begin{align}
     \vert \Pi_3\rangle\longrightarrow\Big(&\frac{\hat{h}_{d_1}^{\dag } \hat{h}_{u_2}^{\dag } \sin \left(\theta\right)}{\sqrt{2}}+\frac{\hat{h}_{d_1}^{\dag } \hat{h}_{u_2}^{\dag } \sin \left(\theta\right) \cos \left(2 \theta
   _1\right)}{\sqrt{2}}\nonumber\\
   &-\frac{\hat{h}_{d_1}^{\dag } \hat{v}_{u_2}^{\dag } \cos \left(\theta\right)}{2 \sqrt{2}}-\frac{\hat{h}_{d_1}^{\dag } \hat{v}_{u_2}^{\dag } \cos \left(3 \theta\right)}{2
   \sqrt{2}}\nonumber\\
   &+\frac{\hat{h}_{u_1}^{\dag } \hat{h}_{d_2}^{\dag } \sin \left(\theta\right)}{\sqrt{2}}+\frac{\hat{h}_{u_1}^{\dag } \hat{h}_{d_2}^{\dag } \sin \left(\theta\right) \cos \left(2 \theta
   _1\right)}{\sqrt{2}}\nonumber\\
   &-\frac{\hat{h}_{u_1}^{\dag } \hat{v}_{d_2}^{\dag } \cos \left(\theta\right)}{2 \sqrt{2}}-\frac{\hat{h}_{u_1}^{\dag } \hat{v}_{d_2}^{\dag } \cos \left(3 \theta\right)}{2
   \sqrt{2}}\nonumber\\
   &+\frac{\hat{v}_{d_1}^{\dag } \hat{v}_{u_2}^{\dag } \sin \left(\theta\right)}{\sqrt{2}}+\frac{\hat{v}_{u_1}^{\dag } \hat{v}_{d_2}^{\dag } \sin \left(\theta\right)}{\sqrt{2}}\Big)\otimes\hat{g}^{\dag}_1\hat{g}^{\dag}_2\ket{0},\label{41}
\end{align}
\begin{align}
     \vert \Pi_4\rangle\longrightarrow\Big(&\frac{\hat{h}_{d_1}^{\dag } \hat{h}_{u_2}^{\dag } \cos \left(\theta\right)}{2 \sqrt{2}}+\frac{\hat{h}_{d_1}^{\dag } \hat{h}_{u_2}^{\dag } \cos \left(3 \theta\right)}{2 \sqrt{2}}\nonumber\\
     &+\frac{\hat{h}_{d_1}^{\dag } \hat{v}_{u_2}^{\dag }
   \sin \left(\theta\right)}{2 \sqrt{2}}+\frac{\hat{h}_{d_1}^{\dag } \hat{v}_{u_2}^{\dag } \sin \left(3 \theta\right)}{2 \sqrt{2}}\nonumber\\
   &+\frac{\hat{h}_{u_1}^{\dag } \hat{h}_{d_2}^{\dag } \cos \left(\theta
   _1\right)}{2 \sqrt{2}}+\frac{\hat{h}_{u_1}^{\dag } \hat{h}_{d_2}^{\dag } \cos \left(3 \theta\right)}{2 \sqrt{2}}\nonumber\\
   &+\frac{\hat{h}_{u_1}^{\dag } \hat{v}_{d_2}^{\dag } \sin \left(\theta\right)}{2
   \sqrt{2}}+\frac{\hat{h}_{u_1}^{\dag } \hat{v}_{d_2}^{\dag } \sin \left(3 \theta\right)}{2 \sqrt{2}}\nonumber\\
   &+\frac{\hat{h}_{d_2}^{\dag } \hat{v}_{u_1}^{\dag } \sin \left(\theta\right)}{\sqrt{2}}+\frac{\hat{h}_{u_2}^{\dag }
   \hat{v}_{d_1}^{\dag } \sin \left(\theta\right)}{\sqrt{2}}\Big)\otimes\hat{g}^{\dag}_1\hat{g}^{\dag}_2\ket{0}.\label{42}
   \end{align}
We make the following observations
\begin{itemize}
    \item For states $\vert \Pi_1\rangle$ and $\vert \Pi_2\rangle$, both the photons are always detected in either paths $u_1$ and $u_2$ or paths $d_1$ and $d_2$. Furthermore, between them, we have the following observations (using Eqs. \eqref{39} and \eqref{40}):
    \begin{enumerate}
        \item Terms $\{\hat{h}_{u_1}^{\dag } \hat{h}_{u_2}^{\dag },\hat{h}_{d_1}^{\dag } \hat{h}_{d_2}^{\dag }\}$ are uniquely present in $\vert \Pi_1\rangle$. Thus, both the photons being horizontally polarised in paths $u_1(d_1)$ and $u_2(d_2)$ indicate unambiguous detection of this state.
        \item Terms $\{\hat{h}_{u_1}^{\dag } \hat{v}_{u_2}^{\dag },\hat{h}_{d_1}^{\dag } \hat{v}_{d_2}^{\dag }\}$ are uniquely present in $\vert \Pi_2\rangle$. Thus, horizontally polarised first photon in paths $u_1(d_1)$ and vertically polarised second photon in paths $u_2(d_2)$ indicates unambiguous detection of this state.
  \end{enumerate}
    \item For states $\vert \Pi_3\rangle$ and $\vert \Pi_4\rangle$, both the photons are always simultaneously detected in both the paths u and d. Furthermore, we have the following observations (see Eqs. \eqref{41} and \eqref{42}):
    \begin{enumerate}
        \item Terms $\{\hat{v}_{u_1}^{\dag } \hat{v}_{d_2}^{\dag },\hat{v}_{d_1}^{\dag } \hat{v}_{u_2}^{\dag }\}$ are uniquely present in $\vert \Pi_3\rangle$. Thus, both the photons being vertically polarised in paths $u_1(d_1)$ and $d_2(u_2)$ indicate unambiguous detection of this state.
  \item Terms $\{\hat{v}_{u_1}^{\dag } \hat{h}_{d_2}^{\dag },\hat{v}_{d_1}^{\dag } \hat{h}_{u_2}^{\dag }\}$  are uniquely present in $\vert \Pi_4\rangle$. Thus vertically polarised first photon in paths $u_1(d_1)$ and horizontally polarised second photon in paths $d_2(u_2)$ indicates unambiguous detection of this state.
    \end{enumerate}
\end{itemize}
There will also be some detection events that will render the discrimination of the Bell-like states ambiguous. Thus, the success probability, unlike Bell state discrimination, will not be $100\%$. If all four Bell-like states are provided with equal probability, then the success probability can be calculated as
\begin{align}
   P_{succ}&=\frac{1}{4}(\frac{\sin^2(\theta)}{2}+\frac{\sin^2(\theta)}{2})+\frac{1}{4}(\frac{\sin^2(\theta)}{2}+\frac{\sin^2(\theta)}{2}),\nonumber\\
    &+\frac{1}{4}(\frac{\cos^2(\theta)}{2}+\frac{\cos^2(\theta)}{2})+\frac{1}{4}(\frac{\cos^2(\theta)}{2}+\frac{\cos^2(\theta)}{2})\nonumber\\
    &=\frac{1}{2}.
\end{align}
Thus, the success probability of unambiguous discrimination is essentially $50\%$.

We would again like to stress the point that the aforesaid protocol may not be the optimal one, but before proceeding forward, we would like to make some comments about it. We would heuristically try to argue that this success probability might be optimal. As discussed earlier, the scheme for optimal discrimination for Bell-like states in Ref. \cite{PhysRevA.106.023706}, can discuss two Bell-like states unambiguously with $25\%$ success probability. It can either distinguish between $\ket{\psi_1}$ and $\ket{\psi_2}$ or $\ket{\psi_3}$ and $\ket{\psi_4}$ but not all four together in Eqs. \eqref{1}-\eqref{4}. If we have access to extra degrees of freedom, exactly this restriction is lifted. For hyperentangled Bell-like states in Eqs.~(\ref{Belllikehypero})-(\ref{388}), we observe the following things:
\begin{itemize}
    \item We cannot make two different degrees of freedom interact/interfere using linear optical elements only, i.e., transformations like
    $$\hat{h}^{\dag}_1\longrightarrow\alpha \hat{h}^{\dag}_1+\beta \hat{p}^{\dag}_1,$$ are not possible.
    \item The same maximally entangled auxiliary state in OAM degrees of freedom of two photons is shared for all Bell-like states in polarisation degrees of freedom. 
\end{itemize}
Hence, doing some operation on OAM degrees of freedom won't generate any new detection signature for discrimination of Bell-like states in polarisation DOF. Thus, the only way it can assist in state discrimination is by using ancillary OAM degrees of freedom to introduce different paths for the pairs $\{\ket{\psi_1},\ket{\psi_2}\}$ and $\{\ket{\psi_3},\ket{\psi_4}\}$.  As we already know, two Bell-like states $\{\ket{\psi_1},\ket{\psi_2}\}$ can be distinguished with $25\%$ success probability. The other pair of states $\{\ket{\psi_3},\ket{\psi_4}\}$ gives similar kinds of detection signatures. Using auxiliary entanglement in other degrees of freedom, these two pairs are in some sense spatially separated. Now both of these pairs can be distinguished by their $25\%$ success probability, which leads to a total of $50\%$ success probability. Thus, it seems that $50\%$ success probability is the optimal value for hyperentanglement-assisted Bell-like state discrimination. We again want to highlight the fact that this is just a heuristic argument, and optimal success probability may be more than $50\%$. As the success probability for hyperentanglement-assisted Bell state discrimination is $100\%$ \cite{PhysRevA.68.042313, PhysRevA.58.R2623, PhysRevLett.118.050501,barreiro2008beating}, this question of optimality never arises. Only in this case of ancilla-assisted Bell state discrimination \cite{PhysRevA.84.042331}, where success probability is $75\%$, the question of optimality is addressed in \cite{PhysRevA.98.042323}. The formalism there is quite involved and requires complex numerical optimisation techniques. Thus, in order to tackle the optimisation problem in full generality, in our case it will be much more involved, as instead of ancilla we have entanglement in additional degrees of freedom of the same photon, and they can't be mixed together as opposed to ancillary photons.

We have discussed three possible combinations of system degrees of freedom and ancilla degrees of freedom. One might ask the question of what about the remaining three possible cases for system qubits and ancilla qubits, namely (i) Bell-like states in OAM degrees of freedom while the ancilla is maximally entangled in polarization degrees of freedom of the two photons, (ii) Bell-like states in momentum degrees of freedom while the ancilla is maximally entangled in OAM degrees of freedom of the two photons, (iii) Bell-like states in OAM degrees of freedom while the ancilla is maximally entangled in momentum degrees of freedom of the two photons. We do think that in these three cases we will also get the same results. Our goal is to compare the Bell-like state discrimination schemes to the corresponding scheme for discrimination of Bell states. Three cases considered here are sufficient for this purpose. They highlight the interesting fact that, due to the asymmetry of Bell-like states, although using hyperentanglement, the success probability for their discrimination can be doubled as compared to the case when no ancillary resources are used; it still only reaches up to $50\%$ and not $100\%$, which is the case for Bell states.

\section{Using correlation in time DOF for Bell-like state discrimination}\label{sec5}
\begin{figure}[h!]
    \centering
    \includegraphics[width=\linewidth]{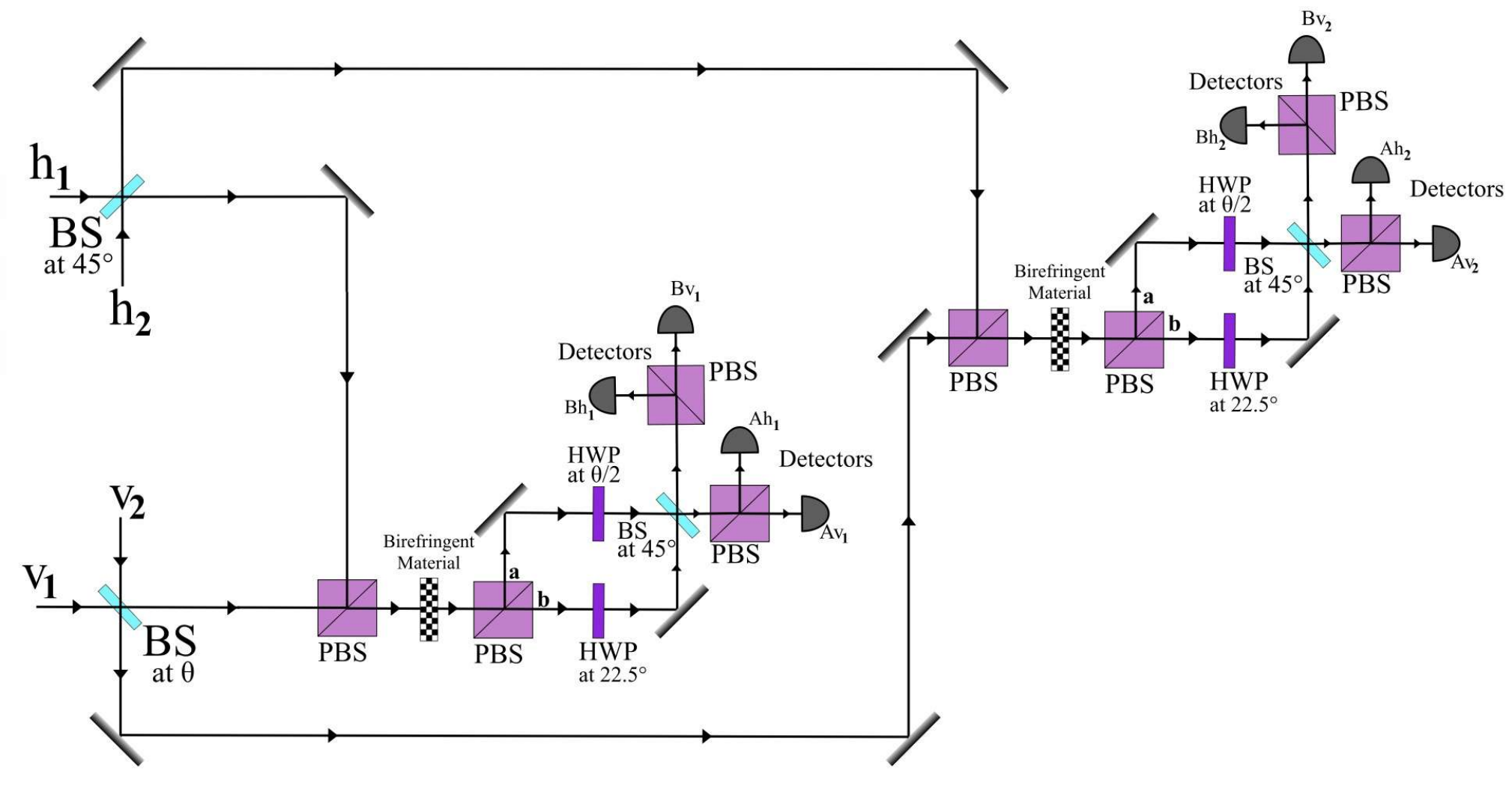}
    \caption{Circuit for a Bell-like state analyzer where correlation in the time DOF of the photon pair is used.}
    \label{fig:5a}
\end{figure}

It is well known that the photon pair coming out of a spontaneous down-conversion source is strongly correlated in time. This correlation can be used to distinguish Bell-like states. It was first proposed by Kwiat and Weinfurter \cite{PhysRevA.58.R2623} for Bell state analysis. We try to do the same for Bell-like states. The circuit is shown in Fig.(\ref{fig:5a}). The philosophy is still the same as in the previous section. As Bell-like states are asymmetric in their coefficients in order to cancel some terms to generate distinct detection signatures we have to introduce some asymmetry in the circuit as well. 

We consider four Bell-like states in polarization degrees of freedom. Mathematically they are given as
\begin{align}
&\vert \psi_1\rangle=(\sin(\theta)\hat{h}^{\dag}_1\hat{v}^{\dag}_2+\cos(\theta)\hat{v}^{\dag}_1\hat{h}^{\dag}_2)\ket{0},\\
            &\vert \psi_2\rangle=(\cos(\theta)\hat{h}^{\dag}_1\hat{v}^{\dag}_2-\sin(\theta)\hat{v}^{\dag}_1\hat{h}^{\dag}_2)\ket{0},\\
            &\vert \psi_3\rangle=(\sin(\theta)\hat{h}^{\dag}_1\hat{h}^{\dag}_2+\cos(\theta)\hat{v}^{\dag}_1\hat{v}^{\dag}_2)\ket{0},\\
            &\vert \psi_4\rangle=(\cos(\theta)\hat{h}^{\dag}_1\hat{h}^{\dag}_2-\sin(\theta)\hat{v}^{\dag}_1\hat{v}^{\dag}_2)\ket{0},
        \end{align}

As is clear from the circuit, horizontally polarised photons are passed through a balanced beam splitter while vertically polarised photons are passed through an unbalanced beam splitter. The output of both beam splitters is then passed through birefringent crystals. These crystals introduce a delay in the passing photon depending on its polarization. A horizontally polarised photon picks up a delay of, say, time $t_h$ while the vertically polarised photon is delayed by time $t_v$. Mathematically, it can be written as:
\begin{align}
    \hat{h}^{\dag}_1&\longrightarrow\hat{h}^{\dag}_1(t_h),\qquad\hat{h}^{\dag}_2\longrightarrow\hat{h}^{\dag}_2(t_h)\\
    \hat{v}^{\dag}_1&\longrightarrow\hat{v}^{\dag}_1(t_v)\qquad~~\hat{v}^{\dag}_2\longrightarrow\hat{v}^{\dag}_2(t_v)
\end{align}
The amount of delay depends on the length of the crystal. We have to bear two things in mind while introducing this delay:
\begin{itemize}
    \item Both the delay times $t_h$ and $t_v$ should be greater than the coherence time of the entangled photon pair. This ensures that both the photons (horizontally and vertically polarised) can be distinguished by their arrival times $t_h$ and $t_v$ at the detectors.
    \item Both of these delay times should be less than the coherence time of the pump photon. This ensures that the photon pair as a whole cannot be distinguished by their times of creation \textit{i.e.} terms like $\hat{x}^{\dag}_1(t_h)\hat{x}^{\dag}_2(t_h)$ and $\hat{x}^{\dag}_1(t_v)\hat{x}^{\dag}_2(t_v)$ are indistinguishable from each other hence will be replaced by  $\hat{x}^{\dag}_1\hat{x}^{\dag}_2$ where $\hat{x}^{\dag}\in\{\hat{h}^{\dag},\hat{v}^{\dag}\}$.
\end{itemize}
The horizontally polarised photons then pass through an HWP aligned at an angle $\frac{\theta}{2}$ while vertically polarised photons are passed through an HWP aligned at the angle of $22.5^{\circ}$. Before the PBS near detectors, the Bell-like states transform as
\begin{widetext}
\begin{align}
            \vert \psi_1\rangle
            &\longrightarrow\nonumber\\&\Big(\frac{1}{4} \hat{h}^{\dagger}_{a_1}(t_h) \hat{h}^{\dagger}_{a_1}(t_v) \cos (\theta)
-\frac{1}{4} \hat{h}^{\dagger}_{a_2}(t_h) \hat{h}^{\dagger}_{a_1}(t_v) \cos (\theta) \cos (2 \theta)
-\frac{1}{4} \hat{h}^{\dagger}_{a_2}(t_h) \hat{h}^{\dagger}_{a_2}(t_v) \sin (2 \theta) \cos (\theta)
-\frac{1}{4} \hat{h}^{\dagger}_{a_1}(t_h) \hat{h}^{\dagger}_{b_1}(t_v) \cos (\theta)\nonumber\\
&+\frac{1}{4} \hat{h}^{\dagger}_{a_2}(t_h) \hat{h}^{\dagger}_{b_1}(t_v) \cos (\theta) \cos (2 \theta)
+\frac{1}{4} \hat{h}^{\dagger}_{a_2}(t_h) \hat{h}^{\dagger}_{b_2}(t_v) \sin (2 \theta) \cos (\theta)
-\frac{1}{4} \hat{h}^{\dagger}_{a_1}(t_h) \hat{v}^{\dagger}_{a_1}(t_v) \cos (\theta)
+\frac{1}{4} \hat{h}^{\dagger}_{a_2}(t_h) \hat{v}^{\dagger}_{a_1}(t_v) \cos (\theta) \cos (2 \theta)\nonumber\\
&+\frac{1}{4} \hat{h}^{\dagger}_{a_2}(t_h) \hat{v}^{\dagger}_{a_2}(t_v) \sin (2 \theta) \cos (\theta)
+\frac{1}{4} \hat{h}^{\dagger}_{a_1}(t_h) \hat{v}^{\dagger}_{b_1}(t_v) \cos (\theta)
-\frac{1}{4} \hat{h}^{\dagger}_{a_2}(t_h) \hat{v}^{\dagger}_{b_1}(t_v) \cos (\theta) \cos (2 \theta)
-\frac{1}{4} \hat{h}^{\dagger}_{a_2}(t_h) \hat{v}^{\dagger}_{b_2}(t_v) \sin (2 \theta) \cos (\theta)\nonumber\\
&+\frac{1}{4} \hat{h}^{\dagger}_{a_1}(t_v) \hat{h}^{\dagger}_{b_1}(t_h) \cos (\theta)
-\frac{1}{4} \hat{h}^{\dagger}_{a_1}(t_v) \hat{h}^{\dagger}_{b_2}(t_h) \cos (\theta) \cos (2 \theta)
-\frac{1}{4} \hat{h}^{\dagger}_{a_2}(t_v) \hat{h}^{\dagger}_{b_2}(t_h) \sin (2 \theta) \cos (\theta)
+\frac{1}{4} \hat{h}^{\dagger}_{a_1}(t_v) \hat{v}^{\dagger}_{a_1}(t_h) \sin (\theta)\nonumber\\
&-\frac{1}{4} \hat{h}^{\dagger}_{a_2}(t_v) \hat{v}^{\dagger}_{a_2}(t_h) \sin (\theta) \sin (2 \theta)
-\frac{1}{4} \hat{h}^{\dagger}_{a_1}(t_v) \hat{v}^{\dagger}_{a_2}(t_h) \sin (\theta) \cos (2 \theta)
+\frac{1}{4} \hat{h}^{\dagger}_{a_1}(t_v) \hat{v}^{\dagger}_{b_1}(t_h) \sin (\theta)
-\frac{1}{4} \hat{h}^{\dagger}_{a_2}(t_v) \hat{v}^{\dagger}_{b_2}(t_h) \sin (\theta) \sin (2 \theta)\nonumber\\
&-\frac{1}{4} \hat{h}^{\dagger}_{a_1}(t_v) \hat{v}^{\dagger}_{b_2}(t_h) \sin (\theta) \cos (2 \theta)
-\frac{1}{4} \hat{h}^{\dagger}_{b_1}(t_h) \hat{h}^{\dagger}_{b_1}(t_v) \cos (\theta)
+\frac{1}{4} \hat{h}^{\dagger}_{b_2}(t_h) \hat{h}^{\dagger}_{b_1}(t_v) \cos (\theta) \cos (2 \theta)
+\frac{1}{4} \hat{h}^{\dagger}_{b_2}(t_h) \hat{h}^{\dagger}_{b_2}(t_v) \sin (2 \theta) \cos (\theta)\nonumber\\
&-\frac{1}{4} \hat{h}^{\dagger}_{b_1}(t_h) \hat{v}^{\dagger}_{a_1}(t_v) \cos (\theta)
+\frac{1}{4} \hat{h}^{\dagger}_{b_2}(t_h) \hat{v}^{\dagger}_{a_1}(t_v) \cos (\theta) \cos (2 \theta)
+\frac{1}{4} \hat{h}^{\dagger}_{b_2}(t_h) \hat{v}^{\dagger}_{a_2}(t_v) \sin (2 \theta) \cos (\theta)
+\frac{1}{4} \hat{h}^{\dagger}_{b_1}(t_h) \hat{v}^{\dagger}_{b_1}(t_v) \cos (\theta)\nonumber\\
&-\frac{1}{4} \hat{h}^{\dagger}_{b_2}(t_h) \hat{v}^{\dagger}_{b_1}(t_v) \cos (\theta) \cos (2 \theta)
-\frac{1}{4} \hat{h}^{\dagger}_{b_2}(t_h) \hat{v}^{\dagger}_{b_2}(t_v) \sin (2 \theta) \cos (\theta)
-\frac{1}{4} \hat{h}^{\dagger}_{b_1}(t_v) \hat{v}^{\dagger}_{a_1}(t_h) \sin (\theta)
+\frac{1}{4} \hat{h}^{\dagger}_{b_2}(t_v) \hat{v}^{\dagger}_{a_2}(t_h) \sin (\theta) \sin (2 \theta)\nonumber\\
&+\frac{1}{4} \hat{h}^{\dagger}_{b_1}(t_v) \hat{v}^{\dagger}_{a_2}(t_h) \sin (\theta) \cos (2 \theta)
-\frac{1}{4} \hat{h}^{\dagger}_{b_1}(t_v) \hat{v}^{\dagger}_{b_1}(t_h) \sin (\theta)
+\frac{1}{4} \hat{h}^{\dagger}_{b_2}(t_v) \hat{v}^{\dagger}_{b_2}(t_h) \sin (\theta) \sin (2 \theta)
+\frac{1}{4} \hat{h}^{\dagger}_{b_1}(t_v) \hat{v}^{\dagger}_{b_2}(t_h) \sin (\theta) \cos (2 \theta)\nonumber\\
&-\frac{1}{4} \hat{v}^{\dagger}_{a_1}(t_h) \hat{v}^{\dagger}_{a_1}(t_v) \sin (\theta)
+\frac{1}{4} \hat{v}^{\dagger}_{a_2}(t_h) \hat{v}^{\dagger}_{a_2}(t_v) \sin (\theta) \sin (2 \theta)
+\frac{1}{4} \hat{v}^{\dagger}_{a_2}(t_h) \hat{v}^{\dagger}_{a_1}(t_v) \sin (\theta) \cos (2 \theta)
+\frac{1}{4} \hat{v}^{\dagger}_{a_1}(t_h) \hat{v}^{\dagger}_{b_1}(t_v) \sin (\theta)\nonumber\\
&-\frac{1}{4} \hat{v}^{\dagger}_{a_2}(t_h) \hat{v}^{\dagger}_{b_2}(t_v) \sin (\theta) \sin (2 \theta)
-\frac{1}{4} \hat{v}^{\dagger}_{a_2}(t_h) \hat{v}^{\dagger}_{b_1}(t_v) \sin (\theta) \cos (2 \theta)
-\frac{1}{4} \hat{v}^{\dagger}_{a_1}(t_v) \hat{v}^{\dagger}_{b_1}(t_h) \sin (\theta)
+\frac{1}{4} \hat{v}^{\dagger}_{a_2}(t_v) \hat{v}^{\dagger}_{b_2}(t_h) \sin (\theta) \sin (2 \theta)\nonumber\\
&+\frac{1}{4} \hat{v}^{\dagger}_{a_1}(t_v) \hat{v}^{\dagger}_{b_2}(t_h) \sin (\theta) \cos (2 \theta)
+\frac{1}{4} \hat{v}^{\dagger}_{b_1}(t_h) \hat{v}^{\dagger}_{b_1}(t_v) \sin (\theta)
-\frac{1}{4} \hat{v}^{\dagger}_{b_2}(t_h) \hat{v}^{\dagger}_{b_2}(t_v) \sin (\theta) \sin (2 \theta)
-\frac{1}{4} \hat{v}^{\dagger}_{b_2}(t_h) \hat{v}^{\dagger}_{b_1}(t_v) \sin (\theta) \cos (2 \theta)\Big)\ket{0},
\end{align}
\begin{align}
\vert \psi_2\rangle
            &\longrightarrow\nonumber\\
&\Big(-\frac{1}{4}\hat{h}^{\dagger}_{a_1}(t_h)\hat{h}^{\dagger}_{a_2}(t_v)\cos (\theta)
-\frac{1}{4}\hat{h}^{\dagger}_{a_2}(t_h)\hat{h}^{\dagger}_{a_2}(t_v)\cos (\theta)\cos (2\theta)
+\frac{1}{4}\hat{h}^{\dagger}_{a_2}(t_h)\hat{h}^{\dagger}_{a_1}(t_v)\sin (2\theta)\cos (\theta)+\frac{1}{4}\hat{h}^{\dagger}_{a_1}(t_h)\hat{h}^{\dagger}_{b_2}(t_v)\cos (\theta)\nonumber
\\
&
+\frac{1}{4}\hat{h}^{\dagger}_{a_2}(t_h)\hat{h}^{\dagger}_{b_2}(t_v)\cos (\theta)\cos (2\theta)
-\frac{1}{4}\hat{h}^{\dagger}_{a_2}(t_h)\hat{h}^{\dagger}_{b_1}(t_v)\sin (2\theta)\cos (\theta)+\frac{1}{4}\hat{h}^{\dagger}_{a_1}(t_h)\hat{v}^{\dagger}_{a_2}(t_v)\cos (\theta)
+\frac{1}{4}\hat{h}^{\dagger}_{a_2}(t_h)\hat{v}^{\dagger}_{a_2}(t_v)\cos (\theta)\cos (2\theta)\nonumber
\\
&
-\frac{1}{4}\hat{h}^{\dagger}_{a_2}(t_h)\hat{v}^{\dagger}_{a_1}(t_v)\sin (2\theta)\cos (\theta)-\frac{1}{4}\hat{h}^{\dagger}_{a_1}(t_h)\hat{v}^{\dagger}_{b_2}(t_v)\cos (\theta)
-\frac{1}{4}\hat{h}^{\dagger}_{a_2}(t_h)\hat{v}^{\dagger}_{b_2}(t_v)\cos (\theta)\cos (2\theta)
+\frac{1}{4}\hat{h}^{\dagger}_{a_2}(t_h)\hat{v}^{\dagger}_{b_1}(t_v)\sin (2\theta)\cos (\theta)\nonumber
\\
&-\frac{1}{4}\hat{h}^{\dagger}_{a_2}(t_v)\hat{h}^{\dagger}_{b_1}(t_h)\cos (\theta)
-\frac{1}{4}\hat{h}^{\dagger}_{a_2}(t_v)\hat{h}^{\dagger}_{b_2}(t_h)\cos (\theta)\cos (2\theta)
+\frac{1}{4}\hat{h}^{\dagger}_{a_1}(t_v)\hat{h}^{\dagger}_{b_2}(t_h)\sin (2\theta)\cos (\theta)-\frac{1}{4}\hat{h}^{\dagger}_{a_2}(t_v)\hat{v}^{\dagger}_{a_1}(t_h)\sin (\theta)\nonumber
\end{align}
\begin{align}
&
+\frac{1}{4}\hat{h}^{\dagger}_{a_1}(t_v)\hat{v}^{\dagger}_{a_2}(t_h)\sin (\theta)\sin (2\theta)
-\frac{1}{4}\hat{h}^{\dagger}_{a_2}(t_v)\hat{v}^{\dagger}_{a_2}(t_h)\sin (\theta)\cos (2\theta)-\frac{1}{4}\hat{h}^{\dagger}_{a_2}(t_v)\hat{v}^{\dagger}_{b_1}(t_h)\sin (\theta)
+\frac{1}{4}\hat{h}^{\dagger}_{a_1}(t_v)\hat{v}^{\dagger}_{b_2}(t_h)\sin (\theta)\sin (2\theta)\nonumber
\\
&
-\frac{1}{4}\hat{h}^{\dagger}_{a_2}(t_v)\hat{v}^{\dagger}_{b_2}(t_h)\sin (\theta)\cos (2\theta)+\frac{1}{4}\hat{h}^{\dagger}_{b_1}(t_h)\hat{h}^{\dagger}_{b_2}(t_v)\cos (\theta)
+\frac{1}{4}\hat{h}^{\dagger}_{b_2}(t_h)\hat{h}^{\dagger}_{b_2}(t_v)\cos (\theta)\cos (2\theta)
-\frac{1}{4}\hat{h}^{\dagger}_{b_2}(t_h)\hat{h}^{\dagger}_{b_1}(t_v)\sin (2\theta)\cos (\theta)\nonumber
\\
&+\frac{1}{4}\hat{h}^{\dagger}_{b_1}(t_h)\hat{v}^{\dagger}_{a_2}(t_v)\cos (\theta)
+\frac{1}{4}\hat{h}^{\dagger}_{b_2}(t_h)\hat{v}^{\dagger}_{a_2}(t_v)\cos (\theta)\cos (2\theta)
-\frac{1}{4}\hat{h}^{\dagger}_{b_2}(t_h)\hat{v}^{\dagger}_{a_1}(t_v)\sin (2\theta)\cos (\theta)-\frac{1}{4}\hat{h}^{\dagger}_{b_1}(t_h)\hat{v}^{\dagger}_{b_2}(t_v)\cos (\theta)\nonumber
\\
&
-\frac{1}{4}\hat{h}^{\dagger}_{b_2}(t_h)\hat{v}^{\dagger}_{b_2}(t_v)\cos (\theta)\cos (2\theta)
+\frac{1}{4}\hat{h}^{\dagger}_{b_2}(t_h)\hat{v}^{\dagger}_{b_1}(t_v)\sin (2\theta)\cos (\theta)+\frac{1}{4}\hat{h}^{\dagger}_{b_2}(t_v)\hat{v}^{\dagger}_{a_1}(t_h)\sin (\theta)
-\frac{1}{4}\hat{h}^{\dagger}_{b_1}(t_v)\hat{v}^{\dagger}_{a_2}(t_h)\sin (\theta)\sin (2\theta)\nonumber
\\
&
+\frac{1}{4}\hat{h}^{\dagger}_{b_2}(t_v)\hat{v}^{\dagger}_{a_2}(t_h)\sin (\theta)\cos (2\theta)+\frac{1}{4}\hat{h}^{\dagger}_{b_2}(t_v)\hat{v}^{\dagger}_{b_1}(t_h)\sin (\theta)
-\frac{1}{4}\hat{h}^{\dagger}_{b_1}(t_v)\hat{v}^{\dagger}_{b_2}(t_h)\sin (\theta)\sin (2\theta)
+\frac{1}{4}\hat{h}^{\dagger}_{b_2}(t_v)\hat{v}^{\dagger}_{b_2}(t_h)\sin (\theta)\cos (2\theta)\nonumber
\\
&-\frac{1}{4}\hat{v}^{\dagger}_{a_2}(t_h)\hat{v}^{\dagger}_{a_1}(t_v)\sin (\theta)\sin (2\theta)
+\frac{1}{4}\hat{v}^{\dagger}_{a_1}(t_h)\hat{v}^{\dagger}_{a_2}(t_v)\sin (\theta)
+\frac{1}{4}\hat{v}^{\dagger}_{a_2}(t_h)\hat{v}^{\dagger}_{a_2}(t_v)\sin (\theta)\cos (2\theta)+\frac{1}{4}\hat{v}^{\dagger}_{a_2}(t_h)\hat{v}^{\dagger}_{b_1}(t_v)\sin (\theta)\sin (2\theta)\nonumber
\\
&
-\frac{1}{4}\hat{v}^{\dagger}_{a_1}(t_h)\hat{v}^{\dagger}_{b_2}(t_v)\sin (\theta)
-\frac{1}{4}\hat{v}^{\dagger}_{a_2}(t_h)\hat{v}^{\dagger}_{b_2}(t_v)\sin (\theta)\cos (2\theta)+\frac{1}{4}\hat{v}^{\dagger}_{a_2}(t_v)\hat{v}^{\dagger}_{b_1}(t_h)\sin (\theta)
-\frac{1}{4}\hat{v}^{\dagger}_{a_1}(t_v)\hat{v}^{\dagger}_{b_2}(t_h)\sin (\theta)\sin (2\theta)\nonumber
\\
&
+\frac{1}{4}\hat{v}^{\dagger}_{a_2}(t_v)\hat{v}^{\dagger}_{b_2}(t_h)\sin (\theta)\cos (2\theta)+\frac{1}{4}\hat{v}^{\dagger}_{b_2}(t_h)\hat{v}^{\dagger}_{b_1}(t_v)\sin (\theta)\sin (2\theta)
-\frac{1}{4}\hat{v}^{\dagger}_{b_1}(t_h)\hat{v}^{\dagger}_{b_2}(t_v)\sin (\theta)
-\frac{1}{4}\hat{v}^{\dagger}_{b_2}(t_h)\hat{v}^{\dagger}_{b_2}(t_v)\sin (\theta)\cos (2\theta)\Big)\ket{0}
\end{align}
\begin{align}
          \vert \psi_3\rangle&\longrightarrow\qquad\qquad\qquad\qquad\qquad\qquad\qquad\qquad\qquad\qquad\qquad\qquad\qquad\qquad\qquad\qquad\qquad\qquad\qquad\qquad\qquad\qquad\qquad\qquad\qquad\qquad\qquad\qquad\qquad\qquad\qquad\qquad\qquad\qquad\qquad\qquad\qquad\qquad\qquad\qquad\qquad\qquad\qquad\qquad\qquad\qquad\nonumber\\&\Big(\frac{1}{4}\hat{v}^{\dagger2}_{a_1}\sin^3(\theta)
-\frac{1}{4}\hat{v}^{\dagger2}_{a_2}\sin^3(\theta)
+\frac{1}{4}\hat{v}^{\dagger2}_{b_1}\sin^3(\theta)
-\frac{1}{4}\hat{v}^{\dagger2}_{b_2}\sin^3(\theta)\nonumber
\\
&+\frac{1}{2}\hat{v}^{\dagger}_{a_1}\hat{v}^{\dagger}_{b_1}\sin^3(\theta)
-\frac{1}{2}\hat{v}^{\dagger}_{a_2}\hat{v}^{\dagger}_{b_2}\sin^3(\theta)+\frac{1}{2}\cos(\theta)\hat{h}^{\dagger}_{a_1}\hat{v}^{\dagger}_{a_1}\sin^2(\theta)
+\frac{1}{2}\cos(\theta)\hat{h}^{\dagger}_{b_1}\hat{v}^{\dagger}_{a_1}\sin^2(\theta)\nonumber
\\
&-\frac{1}{2}\cos(\theta)\hat{h}^{\dagger}_{a_2}\hat{v}^{\dagger}_{a_2}\sin^2(\theta)
-\frac{1}{2}\cos(\theta)\hat{h}^{\dagger}_{b_2}\hat{v}^{\dagger}_{a_2}\sin^2(\theta)+\frac{1}{2}\cos(\theta)\hat{h}^{\dagger}_{a_1}\hat{v}^{\dagger}_{b_1}\sin^2(\theta)
+\frac{1}{2}\cos(\theta)\hat{h}^{\dagger}_{b_1}\hat{v}^{\dagger}_{b_1}\sin^2(\theta)\nonumber
\\
&-\frac{1}{2}\cos(\theta)\hat{h}^{\dagger}_{a_2}\hat{v}^{\dagger}_{b_2}\sin^2(\theta)
-\frac{1}{2}\cos(\theta)\hat{h}^{\dagger}_{b_2}\hat{v}^{\dagger}_{b_2}\sin^2(\theta)+\frac{1}{2}\cos^2(\theta)\hat{h}^{\dagger2}_{a_1}\sin(\theta)
-\frac{1}{2}\cos^2(\theta)\hat{h}^{\dagger2}_{a_2}\sin(\theta)\nonumber
\\
&+\frac{1}{2}\cos^2(\theta)\hat{h}^{\dagger2}_{b_1}\sin(\theta)
-\frac{1}{2}\cos^2(\theta)\hat{h}^{\dagger2}_{b_2}\sin(\theta)+\frac{1}{4}\cos^2(\theta)\hat{v}^{\dagger2}_{a_1}\sin(\theta)
-\frac{1}{4}\cos^2(\theta)\hat{v}^{\dagger2}_{a_2}\sin(\theta)\nonumber
\\
&+\frac{1}{4}\cos^2(\theta)\hat{v}^{\dagger2}_{b_1}\sin(\theta)
-\frac{1}{4}\cos^2(\theta)\hat{v}^{\dagger2}_{b_2}\sin(\theta)-\frac{1}{2}\cos^2(\theta)\hat{h}^{\dagger}_{a_1}\hat{v}^{\dagger}_{a_1}\sin(\theta)
+\frac{1}{2}\cos^2(\theta)\hat{h}^{\dagger}_{b_1}\hat{v}^{\dagger}_{a_1}\sin(\theta)\nonumber
\\
&+\frac{1}{2}\cos^2(\theta)\hat{h}^{\dagger}_{a_2}\hat{v}^{\dagger}_{a_2}\sin(\theta)
-\frac{1}{2}\cos^2(\theta)\hat{h}^{\dagger}_{b_2}\hat{v}^{\dagger}_{a_2}\sin(\theta)+\frac{1}{2}\cos^2(\theta)\hat{h}^{\dagger}_{a_1}\hat{v}^{\dagger}_{b_1}\sin(\theta)
-\frac{1}{2}\cos^2(\theta)\hat{h}^{\dagger}_{b_1}\hat{v}^{\dagger}_{b_1}\sin(\theta)\nonumber
\\
&-\frac{1}{2}\cos^2(\theta)\hat{v}^{\dagger}_{a_1}\hat{v}^{\dagger}_{b_1}\sin(\theta)
-\frac{1}{2}\cos^2(\theta)\hat{h}^{\dagger}_{a_2}\hat{v}^{\dagger}_{b_2}\sin(\theta)+\frac{1}{2}\cos^2(\theta)\hat{h}^{\dagger}_{b_2}\hat{v}^{\dagger}_{b_2}\sin(\theta)
+\frac{1}{2}\cos^2(\theta)\hat{v}^{\dagger}_{a_2}\hat{v}^{\dagger}_{b_2}\sin(\theta)\nonumber
\\
&-\frac{1}{4}\cos(\theta)\cos(2\theta)\hat{h}^{\dagger}_{a_1}\hat{h}^{\dagger}_{a_2}
+\frac{1}{4}\cos(\theta)\cos(2\theta)\hat{h}^{\dagger}_{a_2}\hat{h}^{\dagger}_{b_1}+\frac{1}{4}\cos(\theta)\cos(2\theta)\hat{h}^{\dagger}_{a_1}\hat{h}^{\dagger}_{b_2}
-\frac{1}{4}\cos(\theta)\cos(2\theta)\hat{h}^{\dagger}_{b_1}\hat{h}^{\dagger}_{b_2}\nonumber
\\
&+\frac{1}{4}\cos(\theta)\cos(2\theta)\hat{h}^{\dagger}_{a_2}\hat{v}^{\dagger}_{a_1}
-\frac{1}{4}\cos(\theta)\cos(2\theta)\hat{h}^{\dagger}_{b_2}\hat{v}^{\dagger}_{a_1}+\frac{1}{4}\cos(\theta)\cos(2\theta)\hat{h}^{\dagger}_{a_1}\hat{v}^{\dagger}_{a_2}
-\frac{1}{4}\cos(\theta)\cos(2\theta)\hat{h}^{\dagger}_{b_1}\hat{v}^{\dagger}_{a_2}\nonumber\\
&-\frac{1}{4}\cos(\theta)\cos(2\theta)\hat{v}^{\dagger}_{a_1}\hat{v}^{\dagger}_{a_2}
-\frac{1}{4}\cos(\theta)\cos(2\theta)\hat{h}^{\dagger}_{a_2}\hat{v}^{\dagger}_{b_1}+\frac{1}{4}\cos(\theta)\cos(2\theta)\hat{h}^{\dagger}_{b_2}\hat{v}^{\dagger}_{b_1}
+\frac{1}{4}\cos(\theta)\cos(2\theta)\hat{v}^{\dagger}_{a_2}\hat{v}^{\dagger}_{b_1}\nonumber
\\
&-\frac{1}{4}\cos(\theta)\cos(2\theta)\hat{h}^{\dagger}_{a_1}\hat{v}^{\dagger}_{b_2}
+\frac{1}{4}\cos(\theta)\cos(2\theta)\hat{h}^{\dagger}_{b_1}\hat{v}^{\dagger}_{b_2}+\frac{1}{4}\cos(\theta)\cos(2\theta)\hat{v}^{\dagger}_{a_1}\hat{v}^{\dagger}_{b_2}
-\frac{1}{4}\cos(\theta)\cos(2\theta)\hat{v}^{\dagger}_{b_1}\hat{v}^{\dagger}_{b_2}\Big)\ket{0},
\end{align}
\begin{align}
                       \vert \psi_4\rangle
            &\longrightarrow\qquad\qquad\qquad\qquad\qquad\qquad\qquad\qquad\qquad\qquad\qquad\qquad\qquad\qquad\qquad\qquad\qquad\qquad\qquad\qquad\qquad\qquad\qquad\qquad\qquad\qquad\qquad\qquad\qquad\qquad\qquad\qquad\qquad\qquad\qquad\qquad\qquad\qquad\qquad\qquad\qquad\qquad\qquad\qquad\qquad\qquad\nonumber\\&\Big(\frac{1}{4}\hat{h}^{\dagger2}_{a_1}\cos^3(\theta)
-\frac{1}{4}\hat{h}^{\dagger2}_{a_2}\cos^3(\theta)
+\frac{1}{4}\hat{h}^{\dagger2}_{b_1}\cos^3(\theta)
-\frac{1}{4}\hat{h}^{\dagger2}_{b_2}\cos^3(\theta)\nonumber
\\
&+\frac{1}{2}\hat{h}^{\dagger}_{a_1}\hat{h}^{\dagger}_{b_1}\cos^3(\theta)
-\frac{1}{2}\hat{h}^{\dagger}_{a_2}\hat{h}^{\dagger}_{b_2}\cos^3(\theta)+\frac{1}{2}\hat{h}^{\dagger}_{a_1}\sin(\theta)\hat{v}^{\dagger}_{a_1}\cos^2(\theta)
+\frac{1}{2}\hat{h}^{\dagger}_{b_1}\sin(\theta)\hat{v}^{\dagger}_{a_1}\cos^2(\theta)\nonumber
\\
&-\frac{1}{2}\hat{h}^{\dagger}_{a_2}\sin(\theta)\hat{v}^{\dagger}_{a_2}\cos^2(\theta)
-\frac{1}{2}\hat{h}^{\dagger}_{b_2}\sin(\theta)\hat{v}^{\dagger}_{a_2}\cos^2(\theta)+\frac{1}{2}\hat{h}^{\dagger}_{a_1}\sin(\theta)\hat{v}^{\dagger}_{b_1}\cos^2(\theta)
+\frac{1}{2}\hat{h}^{\dagger}_{b_1}\sin(\theta)\hat{v}^{\dagger}_{b_1}\cos^2(\theta)\nonumber
\\
&-\frac{1}{2}\hat{h}^{\dagger}_{a_2}\sin(\theta)\hat{v}^{\dagger}_{b_2}\cos^2(\theta)
-\frac{1}{2}\hat{h}^{\dagger}_{b_2}\sin(\theta)\hat{v}^{\dagger}_{b_2}\cos^2(\theta)-\frac{1}{4}\hat{h}^{\dagger2}_{a_1}\sin^2(\theta)\cos(\theta)
+\frac{1}{4}\hat{h}^{\dagger2}_{a_2}\sin^2(\theta)\cos(\theta)\nonumber
\\
&-\frac{1}{4}\hat{h}^{\dagger2}_{b_1}\sin^2(\theta)\cos(\theta)
+\frac{1}{4}\hat{h}^{\dagger2}_{b_2}\sin^2(\theta)\cos(\theta)+\frac{1}{2}\hat{h}^{\dagger}_{a_1}\hat{h}^{\dagger}_{b_1}\sin^2(\theta)\cos(\theta)
-\frac{1}{2}\hat{h}^{\dagger}_{a_2}\hat{h}^{\dagger}_{b_2}\sin^2(\theta)\cos(\theta)\nonumber
\end{align}
\begin{align}
&+\frac{1}{2}\hat{h}^{\dagger}_{a_1}\sin^2(\theta)\hat{v}^{\dagger}_{a_1}\cos(\theta)
-\frac{1}{2}\hat{h}^{\dagger}_{b_1}\sin^2(\theta)\hat{v}^{\dagger}_{a_1}\cos(\theta)-\frac{1}{2}\hat{h}^{\dagger}_{a_2}\sin^2(\theta)\hat{v}^{\dagger}_{a_2}\cos(\theta)
+\frac{1}{2}\hat{h}^{\dagger}_{b_2}\sin^2(\theta)\hat{v}^{\dagger}_{a_2}\cos(\theta)\quad\quad\qquad\qquad\qquad\qquad\nonumber
\\
&-\frac{1}{2}\hat{h}^{\dagger}_{a_1}\sin^2(\theta)\hat{v}^{\dagger}_{b_1}\cos(\theta)
+\frac{1}{2}\hat{h}^{\dagger}_{b_1}\sin^2(\theta)\hat{v}^{\dagger}_{b_1}\cos(\theta)+\sin^2(\theta)\hat{v}^{\dagger}_{a_1}\hat{v}^{\dagger}_{b_1}\cos(\theta)
+\frac{1}{2}\hat{h}^{\dagger}_{a_2}\sin^2(\theta)\hat{v}^{\dagger}_{b_2}\cos(\theta)\nonumber
\\
&-\frac{1}{2}\hat{h}^{\dagger}_{b_2}\sin^2(\theta)\hat{v}^{\dagger}_{b_2}\cos(\theta)
-\sin^2(\theta)\hat{v}^{\dagger}_{a_2}\hat{v}^{\dagger}_{b_2}\cos(\theta)+\frac{1}{4}\cos(2\theta)\hat{h}^{\dagger}_{a_1}\hat{h}^{\dagger}_{a_2}\sin(\theta)
-\frac{1}{4}\cos(2\theta)\hat{h}^{\dagger}_{a_2}\hat{h}^{\dagger}_{b_1}\sin(\theta)\nonumber
\\
&-\frac{1}{4}\cos(2\theta)\hat{h}^{\dagger}_{a_1}\hat{h}^{\dagger}_{b_2}\sin(\theta)
+\frac{1}{4}\cos(2\theta)\hat{h}^{\dagger}_{b_1}\hat{h}^{\dagger}_{b_2}\sin(\theta)-\frac{1}{4}\cos(2\theta)\hat{h}^{\dagger}_{a_2}\sin(\theta)\hat{v}^{\dagger}_{a_1}
+\frac{1}{4}\cos(2\theta)\hat{h}^{\dagger}_{b_2}\sin(\theta)\hat{v}^{\dagger}_{a_1}\nonumber
\\
&-\frac{1}{4}\cos(2\theta)\hat{h}^{\dagger}_{a_1}\sin(\theta)\hat{v}^{\dagger}_{a_2}
+\frac{1}{4}\cos(2\theta)\hat{h}^{\dagger}_{b_1}\sin(\theta)\hat{v}^{\dagger}_{a_2}+\frac{1}{4}\cos(2\theta)\sin(\theta)\hat{v}^{\dagger}_{a_1}\hat{v}^{\dagger}_{a_2}
+\frac{1}{4}\cos(2\theta)\hat{h}^{\dagger}_{a_2}\sin(\theta)\hat{v}^{\dagger}_{b_1}\nonumber
\\
&-\frac{1}{4}\cos(2\theta)\hat{h}^{\dagger}_{b_2}\sin(\theta)\hat{v}^{\dagger}_{b_1}
-\frac{1}{4}\cos(2\theta)\sin(\theta)\hat{v}^{\dagger}_{a_2}\hat{v}^{\dagger}_{b_1}+\frac{1}{4}\cos(2\theta)\hat{h}^{\dagger}_{a_1}\sin(\theta)\hat{v}^{\dagger}_{b_2}
-\frac{1}{4}\cos(2\theta)\hat{h}^{\dagger}_{b_1}\sin(\theta)\hat{v}^{\dagger}_{b_2}\nonumber
\\
&-\frac{1}{4}\cos(2\theta)\sin(\theta)\hat{v}^{\dagger}_{a_1}\hat{v}^{\dagger}_{b_2}
+\frac{1}{4}\cos(2\theta)\sin(\theta)\hat{v}^{\dagger}_{b_1}\hat{v}^{\dagger}_{b_2}\Big)\ket{0}.
\end{align}
\end{widetext}
In the case of the states $\{\vert\psi_1\rangle$, $\vert\psi_2\rangle\}$, as both the photons have different polarization, a time delay of $|t_h-t_v|$ is introduced between them due to birefringent material, but for the states $\{\vert\psi_3\rangle$, $\vert\psi_4\rangle\}$, as both the photons have the same polarization, no time delay is present between the detection of the two photons. This fact renders the pair of states $\{\vert\psi_1\rangle$, $\vert\psi_2\rangle\}$ distinguishable from the pair of states $\{\vert\psi_3\rangle$, $\vert\psi_4\rangle\}$. We then make the following observations:
\begin{enumerate}
    \item Terms \begin{align}
\Big\{&
\hat{h}^{\dagger}_{a_1}(t_h)\hat{h}^{\dagger}_{a_1}(t_v),
\hat{h}^{\dagger}_{a_1}(t_v)\hat{h}^{\dagger}_{b_1}(t_h),
\hat{h}^{\dagger}_{a_1}(t_h)\hat{h}^{\dagger}_{b_1}(t_v),
\hat{h}^{\dagger}_{b_1}(t_h)\hat{h}^{\dagger}_{b_1}(t_v),\nonumber\\
&
\hat{h}^{\dagger}_{a_1}(t_v)\hat{v}^{\dagger}_{a_1}(t_h),
\hat{h}^{\dagger}_{b_1}(t_v)\hat{v}^{\dagger}_{a_1}(t_h),
\hat{h}^{\dagger}_{a_1}(t_h)\hat{v}^{\dagger}_{a_1}(t_v),
\hat{h}^{\dagger}_{b_1}(t_h)\hat{v}^{\dagger}_{a_1}(t_v),\nonumber\\
&
\hat{v}^{\dagger}_{a_1}(t_h)\hat{v}^{\dagger}_{a_1}(t_v),
\hat{h}^{\dagger}_{a_1}(t_v)\hat{v}^{\dagger}_{b_1}(t_h),
\hat{h}^{\dagger}_{b_1}(t_v)\hat{v}^{\dagger}_{b_1}(t_h),
\hat{v}^{\dagger}_{a_1}(t_v)\hat{v}^{\dagger}_{b_1}(t_h),\nonumber\\
&
\hat{h}^{\dagger}_{a_1}(t_h)\hat{v}^{\dagger}_{b_1}(t_v),
\hat{h}^{\dagger}_{b_1}(t_h)\hat{v}^{\dagger}_{b_1}(t_v),
\hat{v}^{\dagger}_{a_1}(t_h)\hat{v}^{\dagger}_{b_1}(t_v),
\hat{v}^{\dagger}_{b_1}(t_h)\hat{v}^{\dagger}_{b_1}(t_v)\nonumber
\Big\}
\end{align} are uniquely present in $\vert\psi_1\rangle$. Thus, detection of 
\begin{itemize}
    \item both photons in detector $A (Ah_1, Av_1)$
    \item or both photons in detector $B (Bh_1, Bv_1)$
    \item or one photon in detector $A (Ah_1, Av_1)$ and another in $B (Bh_1, Bv_1)$
\end{itemize} 
with a delay of $|t_h-t_v|$ between them, indicates the presence of this state.
     \item Terms \begin{align}
\Big\{&
\hat{h}^{\dagger}_{a_1}(t_h)\hat{h}^{\dagger}_{a_2}(t_v),
\hat{h}^{\dagger}_{a_2}(t_v)\hat{h}^{\dagger}_{b_1}(t_h),
\hat{h}^{\dagger}_{a_1}(t_h)\hat{h}^{\dagger}_{b_2}(t_v),
\hat{h}^{\dagger}_{b_1}(t_h)\hat{h}^{\dagger}_{b_2}(t_v),\nonumber\\
&
\hat{h}^{\dagger}_{a_2}(t_v)\hat{v}^{\dagger}_{a_1}(t_h),
\hat{h}^{\dagger}_{b_2}(t_v)\hat{v}^{\dagger}_{a_1}(t_h),
\hat{h}^{\dagger}_{a_1}(t_h)\hat{v}^{\dagger}_{a_2}(t_v),
\hat{h}^{\dagger}_{b_1}(t_h)\hat{v}^{\dagger}_{a_2}(t_v),\nonumber\\
&
\hat{v}^{\dagger}_{a_1}(t_h)\hat{v}^{\dagger}_{a_2}(t_v),
\hat{h}^{\dagger}_{a_2}(t_v)\hat{v}^{\dagger}_{b_1}(t_h),
\hat{h}^{\dagger}_{b_2}(t_v)\hat{v}^{\dagger}_{b_1}(t_h),
\hat{v}^{\dagger}_{a_2}(t_v)\hat{v}^{\dagger}_{b_1}(t_h),\nonumber\\
&
\hat{h}^{\dagger}_{a_1}(t_h)\hat{v}^{\dagger}_{b_2}(t_v),
\hat{h}^{\dagger}_{b_1}(t_h)\hat{v}^{\dagger}_{b_2}(t_v),
\hat{v}^{\dagger}_{a_1}(t_h)\hat{v}^{\dagger}_{b_2}(t_v),
\hat{v}^{\dagger}_{b_1}(t_h)\hat{v}^{\dagger}_{b_2}(t_v)\nonumber
\Big\}
\end{align} are uniquely present in $\vert\psi_2\rangle$. Thus, detection of 
\begin{itemize}
    \item one photon in detector $A (Ah_1, Av_1)$ and the another photon in detector $A (Ah_2, Av_2)$
    \item or one photon in detector $A (Ah_1, Av_1)$ and the another photon in detector $B (Bh_2, Bv_2)$
    \item or one photon in detector $B (Bh_1, Bv_1)$ and the another photon in detector $A (Ah_2, Av_2)$
    \item or one photon in detector $B (Bh_1, Bv_1)$ and the another photon in detector $B (Bh_2, Bv_2)$
\end{itemize} with the delay of $|t_h-t_v|$ between them indicates the presence of this state.
     \item Terms $\left\{
\hat{v}^{\dagger 2}_{a_1},
\hat{v}^{\dagger 2}_{a_2},
\hat{v}^{\dagger 2}_{b_1},
\hat{v}^{\dagger 2}_{b_2}
\right\}$ are uniquely present in $\vert\psi_3\rangle$. Thus, if both photons are vertically polarized and detected simultaneously without any delay in any of the detectors $Av_1, Bv_1, Av_2, Bv_2$ the presence of this state is confirmed. 
\item Terms $\left\{
\hat{h}^{\dagger}_{a_1}\hat{h}^{\dagger}_{b_1},
\hat{h}^{\dagger}_{a_2}\hat{h}^{\dagger}_{b_2}
\right\}$ are uniquely present in $\vert\psi_4\rangle$. Thus, if both photons are horizontally polarised and are detected simultaneously without any delay in either detectors  $Ah_1$ and $Bh_1$ or detectors $Ah_2$ and $Bh_2$, the presence of this state is confirmed. 
\end{enumerate}

The success probability is necessarily not $100\%$. If we consider that all four states are provided to us with equal probability, then the success probability is:
\begin{align}
   P_{succ}=&\frac{1}{4}\Big(\frac{1}{2}\Big)+\frac{1}{4}\Big(\frac{1}{2}\Big)+\frac{1}{4}\Big(\frac{\sin^2(\theta)}{2}\Big)+\frac{1}{4}(\frac{\cos^2(\theta)}{2}),\nonumber\\
    =&\frac{3}{8}.\label{timebinsuccprob}
\end{align}

Evidently, the present scheme, although it provides an unambiguous discrimination of all four Bell-Like states, has a success probability of $37.5\%$ and not $50\%$ as in the previous setting where hyperentanglement was utilized. This is because of the fact that although we are using the measurement of the existence of zero versus non-zero time delay to distinguish among the four Bell-like states, due to their asymmetry, it can only distinguish between $\vert\psi_3\rangle$ and $\vert\psi_4\rangle$ with a success probability of $1/8$, as evident from the expression in Eq. \eqref{timebinsuccprob}. In the case of hyperentanglement, because we have access to extra degrees of freedom of the photon, it is possible to distinguish between the states $\{\vert\psi_1\rangle$, $\vert\psi_2\rangle\}$ and $\{\vert\psi_3\rangle$, $\vert\psi_4\rangle\}$ optimally with $25\%$ success probability for each pair. Hence, the success probability increases to $50\%$ 

Thus, we observe that in case correlations in time are used to assist in the discrimination of all four Bell-like states, the success probability, although higher than $ 25\%$, unlike previous cases, never reaches $50\%$, let alone $100\%$, and only goes up to $37.5\%$. In the case of $\theta=\dfrac{\pi}{4}$, we get back the results of Ref. \cite{PhysRevA.58.R2623} where the success probability of Bell state discrimination using correlations in time is $100\%$.
\section{Bell-like state discrimination using Ancillary qubits}\label{sec6}
\begin{figure}[h!]
    \centering
    \includegraphics[width=\linewidth]{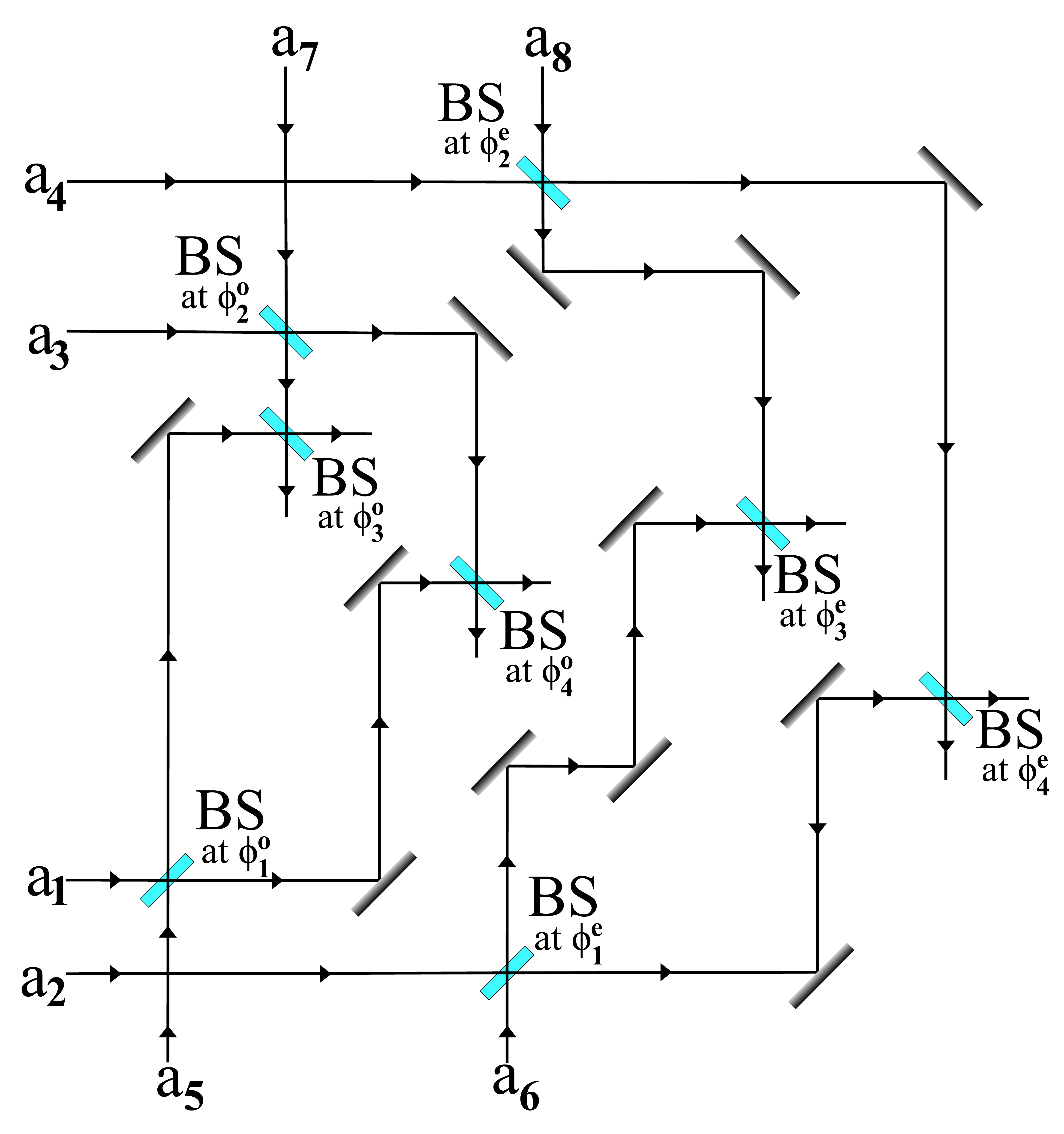}
    \caption{Circuit for the Bell-like state analyzer using ancillary qubits for assistance.}
    \label{fig:7}
\end{figure}

\begin{figure}[h!]
    \centering
    \includegraphics[scale=0.35]{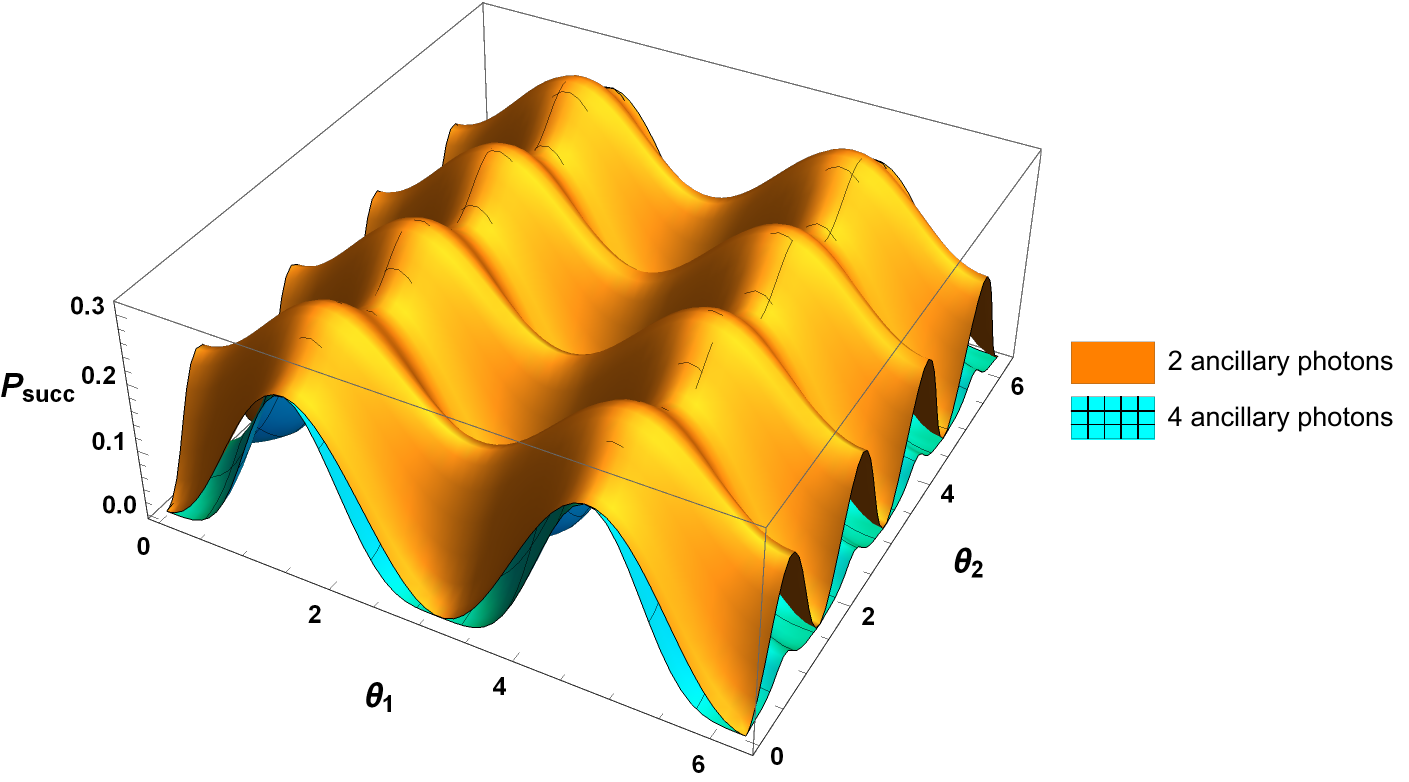}
    \caption{Variation of success probability of Bell-like state discrimination against state parameters -- in the context of using ancillary photon pairs in entangled states.}
    \label{fig:8}
\end{figure}

Aside from using hyperentanglement, extra ancillary photons can come to our aid for Bell-like state analysis. It may be noted here that from the perspective of photon detection, this analysis (where we use extra entangled photon pairs in the ancilla) is more challenging, as we need to probe/process more than one photon simultaneously -- unlike in the previous sections. Grice found that by using two ancillary maximally entangled photons \cite{PhysRevA.84.042331}, the success probability for Bell state discrimination can be enhanced to $75\%$ from $50\%$. Only two Bell states ($\ket{\psi}^+$ and $\ket{\psi}^-$) are distinguishable without using any resource in a linear optical setting. The introduction of ancillary entangled qubits renders the remaining two ($\ket{\phi}^+$ and $\ket{\phi}^-$) also distinguishable (not unambiguously, as there will be some detection signatures that will be common for both states). Thus, the success probability of Bell state discrimination increases from $50\%$ (two states are distinguishable without any ambiguity) to $75\%$ (all four can be distinguished with some ambiguity in ($\ket{\phi}^+$ and $\ket{\phi}^-$). The circuit is the generalization of the circuit considered for Bell state discrimination\cite{Braun_1995,Weinfurter_1994,bouwmeester1997experimental,Bianchi_2026} to account for ancillary photons.  We try to do a similar kind of thing here. We generalize the circuit for optimal discrimination of Bell-like states in \cite{PhysRevA.106.023706} in the same way. The goal here is to see whether this straightforward generalization leads to an increase in success probability and compare it to the improvement for the corresponding case of Bell state discrimination in \cite{PhysRevA.84.042331}. The Bell-like states with ancillary photons are given as
\begin{align}
    \vert \Gamma_1\rangle=&\left(\hat{a}^{\dag}_1 \hat{a}^{\dag}_4 \sin \left(\theta _2\right)+\hat{a}^{\dag}_2 \hat{a}^{\dag}_3 \cos \left(\theta _2\right)\right) \nonumber\\
    &\qquad \qquad \qquad \otimes\left(\hat{a}^{\dag}_5 \hat{a}^{\dag}_7 \sin \left(\theta
   _1\right)+\hat{a}^{\dag}_6 \hat{a}^{\dag}_8 \cos \left(\theta _1\right)\right)\ket{0},\\
   \vert \Gamma_2\rangle=&\left(\hat{a}^{\dag}_1 \hat{a}^{\dag}_4 \cos \left(\theta _2\right)-\hat{a}^{\dag}_2 \hat{a}^{\dag}_3 \sin \left(\theta _2\right)\right)\nonumber\\
    &\qquad \qquad \qquad \otimes \left(\hat{a}^{\dag}_5 \hat{a}^{\dag}_7 \sin \left(\theta
   _1\right)+\hat{a}^{\dag}_6 \hat{a}^{\dag}_8 \cos \left(\theta _1\right)\right)\ket{0},\\
    \vert \Gamma_3\rangle=&\left(\hat{a}^{\dag}_1 \hat{a}^{\dag}_3 \sin \left(\theta _1\right)+\hat{a}^{\dag}_2 \hat{a}^{\dag}_4 \cos \left(\theta _1\right)\right)\nonumber\\
    &\qquad \qquad \qquad \otimes \left(\hat{a}^{\dag}_5 \hat{a}^{\dag}_7 \sin \left(\theta
   _1\right)+\hat{a}^{\dag}_6 \hat{a}^{\dag}_8 \cos \left(\theta _1\right)\right)\ket{0},\\
   \vert \Gamma_4\rangle=&\left(\hat{a}^{\dag}_1 \hat{a}^{\dag}_3 \cos \left(\theta _1\right)-\hat{a}^{\dag}_2 \hat{a}^{\dag}_4 \sin \left(\theta _1\right)\right)\nonumber\\
    &\qquad \qquad \qquad \otimes \left(\hat{a}^{\dag}_5 \hat{a}^{\dag}_7 \sin \left(\theta
   _1\right)+\hat{a}^{\dag}_6 \hat{a}^{\dag}_8 \cos \left(\theta _1\right)\right)\ket{0},
\end{align}
where $\theta_1$ and $\theta_2$ are the state parameters. The circuit for Bell-like state analysis is given in Fig.(\ref{fig:7}).
Here $\phi_1^o=\phi_2^o=\phi_3^o=\phi_4^o=\frac{\pi}{4}$ and  $\phi_1^e=\phi_2^e=\phi_3^e=\phi_4^e=\theta_2$. It turns out that we can discriminate among $\vert\Gamma_1\rangle, \vert\Gamma_2\rangle$ and $\vert\Gamma_4\rangle$ with the success probability of
\begin{equation}
   P_{succ}=\frac{1}{32} \left(-2 \cos \left(4 \theta _2\right)-\cos \left(2 \theta _1\right) \left(\cos \left(4 \theta _2\right)+3\right)+6\right).\label{succprob1anc}
\end{equation}
 Detection signatures can be summarised as:
 \begin{itemize}
     \item If the set of output modes $\{\hat{a}^{\dag}_{1}\hat{a}^{\dag}_{3}\hat{a}^{\dag}_{5}\hat{a}^{\dag}_{7}\}$ and $\{\hat{a}^{\dag}_{2}\hat{a}^{\dag}_{4}\hat{a}^{\dag}_{6}\hat{a}^{\dag}_{8}\}$ each have odd number of photons in total, states $\vert \Gamma_1\rangle$ and $\vert \Gamma_2\rangle$ are present.
     \begin{enumerate}
         \item If the set of output modes $\{\hat{a}^{\dag}_{1}\hat{a}^{\dag}_{2}\hat{a}^{\dag}_{5}\hat{a}^{\dag}_{6}\}$ have even no. of photons in total, we can unambiguously detect $\vert \Gamma_1\rangle$.
         \item If the set of output modes $\{\hat{a}^{\dag}_{1}\hat{a}^{\dag}_{2}\hat{a}^{\dag}_{5}\hat{a}^{\dag}_{6}\}$ has an odd number of photons in total, we can unambiguously detect $\vert \Gamma_2\rangle$.
     \end{enumerate}
     \item If the set of output modes $\{\hat{a}^{\dag}_{1}\hat{a}^{\dag}_{3}\hat{a}^{\dag}_{5}\hat{a}^{\dag}_{7}\}$ and $\{\hat{a}^{\dag}_{2}\hat{a}^{\dag}_{4}\hat{a}^{\dag}_{6}\hat{a}^{\dag}_{8}\}$ each have an even number of photons in total, states $\vert \Gamma_3\rangle$ and $\vert \Gamma_4\rangle$ are present.
     \begin{enumerate}
         \item If two output modes are from the set $\{\hat{a}^{\dag}_{1}, \hat{a}^{\dag}_{3}, \hat{a}^{\dag}_{5}, \hat{a}^{\dag}_{7}\}$ and the other two are from the set $\{\hat{a}^{\dag}_{2}, \hat{a}^{\dag}_{4}, \hat{a}^{\dag}_{6}, \hat{a}^{\dag}_{8}\}$ only then we can uniquely detect $\vert \Gamma_4\rangle$
         \item Furthermore, the set of output modes $\{\hat{a}^{\dag}_{1}\hat{a}^{\dag}_{2}\hat{a}^{\dag}_{3}\hat{a}^{\dag}_{4}\}$ has an odd number of photons in total, and the presence of exactly two photons in any one of the $\{\hat{a}^{\dag}_{1}, \hat{a}^{\dag}_{3}, \hat{a}^{\dag}_{5}, \hat{a}^{\dag}_{7}\}$ is the signature for unambiguous detection of this state.
         \end{enumerate}
 \end{itemize}
A similar thing can be done if we add another 4 extra ancillary entangled photons of the form $\big(\hat{a}^{\dag}_9 \hat{a}^{\dag}_{11} \hat{a}^{\dag}_{13} \hat{a}^{\dag}_{15} \sin
   \left(\theta _1\right)+\hat{a}^{\dag}_{10} \hat{a}^{\dag}_{12} \hat{a}^{\dag}_{14} \hat{a}^{\dag}_{16} \cos \left(\theta
   _1\right)\Big)\ket{0}$.
   The expression for the success probability turns out to be very complex. It is given as follows:
   \begin{widetext}
 \begin{align}
P_{succ}=&\frac{1}{536870912}\Bigg(
67534737
-18271885 \cos(4\theta_2)
+65536 \cos(4\theta_1)\Big(178+81\cos(4\theta_2)\Big)
\nonumber\\
&+\cos(2\theta_1)\Big(
-71794799
+3813747 \cos(4\theta_2)
-262144 \cos^2(\theta_1)\Big(2\cos(8\theta_2)+\cos(12\theta_2)\Big)
\Big)
\nonumber\\
&-\cos^2(\theta_1)\Big(
-3838064 \cos(8\theta_2)
+1070124 \cos(12\theta_2)
-106088 \cos(16\theta_2)
\nonumber\\
&
+196220 \cos(20\theta_2)+137712 \cos(24\theta_2)
-1917 \cos(28\theta_2)+5130 \cos(32\theta_2)
+6075 \cos(36\theta_2)
\Big)
\Bigg).\label{succprob2anc}
 \end{align}
   \end{widetext}
   The variation of success probabilities (of Eqs. \ref{succprob1anc} and \ref{succprob2anc}) with state parameters $\theta_1$ and $\theta_2$ is given in Fig.(\ref{fig:8}).
We observe two things:
\begin{itemize}
    \item The success probability goes beyond $25\%$ for certain values of state parameters $\theta_1$ and $\theta_2$.
    \item Counterintuitively, as we increase the number of ancillary photons, the success probability decreases for given values of state parameters $\theta_1$ and $\theta_2$. It still goes beyond $25\%$ in certain state parameter regimes.
\end{itemize}

An intuitive reason behind this is that as the number of ancillary photons increases, the number of possible detection signatures also increases. Due to a lack of symmetry for Bell-like states, the events corresponding to ambiguous signatures increase at a much faster rate than the terms corresponding to signatures indicating the presence of a particular Bell-like state.

\begin{figure}[h!]
    \centering
    \includegraphics[scale=0.45]{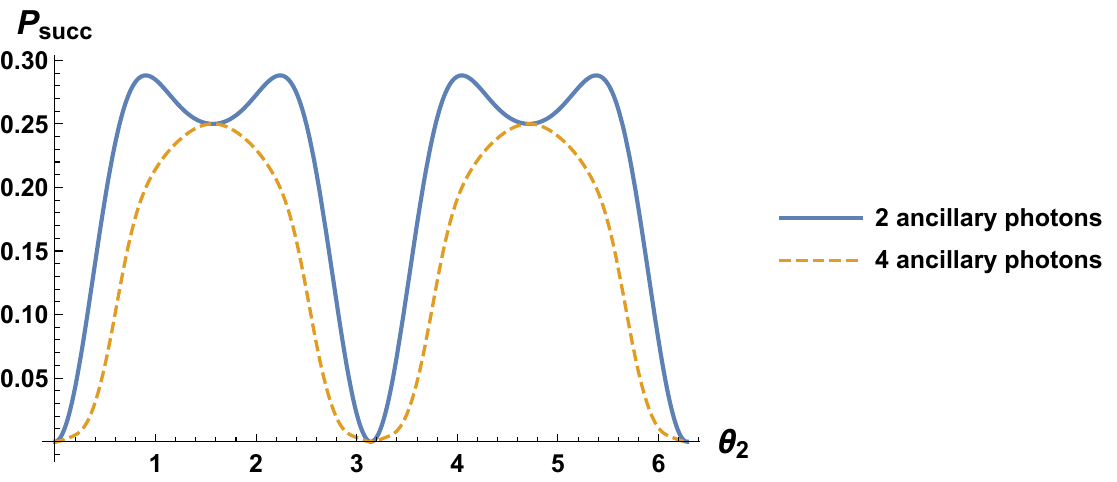}
    \caption{Variation of success probability of Bell-like state discrimination against the state parameter}
    \label{fig:9}
\end{figure}

 If we take $\theta_1=\theta_2$, then we can unambiguously discriminate all four Bell-like states. The detection signatures remain exactly the same, with the extra addition being
 \begin{itemize}
     \item The set of output modes $\{\hat{a}^{\dag}_{1}\hat{a}^{\dag}_{2}\hat{a}^{\dag}_{3}\hat{a}^{\dag}_{4}\}$ has an even number of photons, and the presence of exactly two photons in any one of the modes $\{\hat{a}^{\dag}_{1}, \hat{a}^{\dag}_{3}, \hat{a}^{\dag}_{5}, \hat{a}^{\dag}_{7}\}$ is the signature for unambiguous detection of $\vert \Gamma_3\rangle$.
 \end{itemize}
 The success probability of the Bell-like state discrimination using two ancillary entangled photons can (when ${\theta}_1 = {\theta}_2$) be calculated as
 \begin{equation}
    P_{succ}=\frac{1}{16} \sin ^2\left(\theta _2\right) \left(7 \cos \left(2 \theta _2\right)+\cos \left(4 \theta _2\right)+10\right).
 \end{equation}
Just like we did previously, we can add extra ancillary entangled photons of the same form for assistance and calculate the success probability. The expression for the success probability is again very complicated, as evident from the following:
\begin{widetext}
\begin{align}
P_{succ}=&\frac{1}{536870912}\Bigg(
-21976952 \cos(2\theta_2)
-34846438 \cos(4\theta_2)
-318496 \cos(6\theta_2)
 \nonumber\\
&+16059216 \cos(8\theta_2)+13226464 \cos(10\theta_2)
+7756844 \cos(12\theta_2)
+5213904 \cos(14\theta_2)
 \nonumber\\
&+4028216 \cos(16\theta_2)+2761424 \cos(18\theta_2)
+1584764 \cos(20\theta_2)
+3\Big(
46362914
 \nonumber\\
&
+266848 \cos(22\theta_2)+116752 \cos(24\theta_2)
+58464 \cos(26\theta_2)
+45441 \cos(28\theta_2)
 \nonumber\\
&
+31140 \cos(30\theta_2)+17910 \cos(32\theta_2)
+8100 \cos(34\theta_2)
+2025 \cos(36\theta_2)
\Big)
\Bigg)\sin^2(\theta_2).
\end{align}
\end{widetext}
These are plotted in the Fig.(\ref{fig:9})
We again observe that for a certain value of the state parameter, the success probability can go beyond $25\%$. Also, as we increase the number of ancillary photons, the success probability for a given value of the state parameter decreases. The only difference being that in the latter case we can discriminate all four Bell-like states. Comparing it to Bell-state discrimination using ancillary photons, what we observe is that due to the lack of symmetry, Bell-like states behave very differently. Thus, a much more careful and involved treatment is required for discriminating among them, and straightforward generalization of the optimal scheme for Bell-like state discrimination without using any ancillary resources in Ref. \cite{PhysRevA.106.023706} is not always efficient.

\section{Bell-like state analysis using Non-linear elements}\label{sec7}
Besides using hyperentanglement or extra ancillary photons, we can take the aid of non-linear gadgets with the aim of completely discriminating among Bell states. Four Bell-like states in a polarization basis are given as
 \begin{align}
            &\vert \psi_1\rangle=(\sin(\theta_1)\hat{h}^{\dag}_1\hat{h}^{\dag}_2+\cos(\theta_1)\hat{v}^{\dag}_1\hat{v}^{\dag}_2)\ket{0},\\
            &\vert \psi_2\rangle=(\cos(\theta_1)\hat{h}^{\dag}_1\hat{h}^{\dag}_2-\sin(\theta_1)\hat{v}^{\dag}_1\hat{v}^{\dag}_2)\ket{0},\\
            &\vert \psi_3\rangle=(\sin(\theta_2)\hat{h}^{\dag}_1\hat{v}^{\dag}_2+\cos(\theta_2)\hat{v}^{\dag}_1\hat{h}^{\dag}_2)\ket{0},\\
            &\vert \psi_4\rangle=(\cos(\theta_2)\hat{h}^{\dag}_1\hat{v}^{\dag}_2-\sin(\theta_2)\hat{v}^{\dag}_1\hat{h}^{\dag}_2)\ket{0}.
        \end{align}
\subsection*{Using Sum Frequency Generation}        
\begin{figure}[h!]
    \centering
    \includegraphics[width=\linewidth]{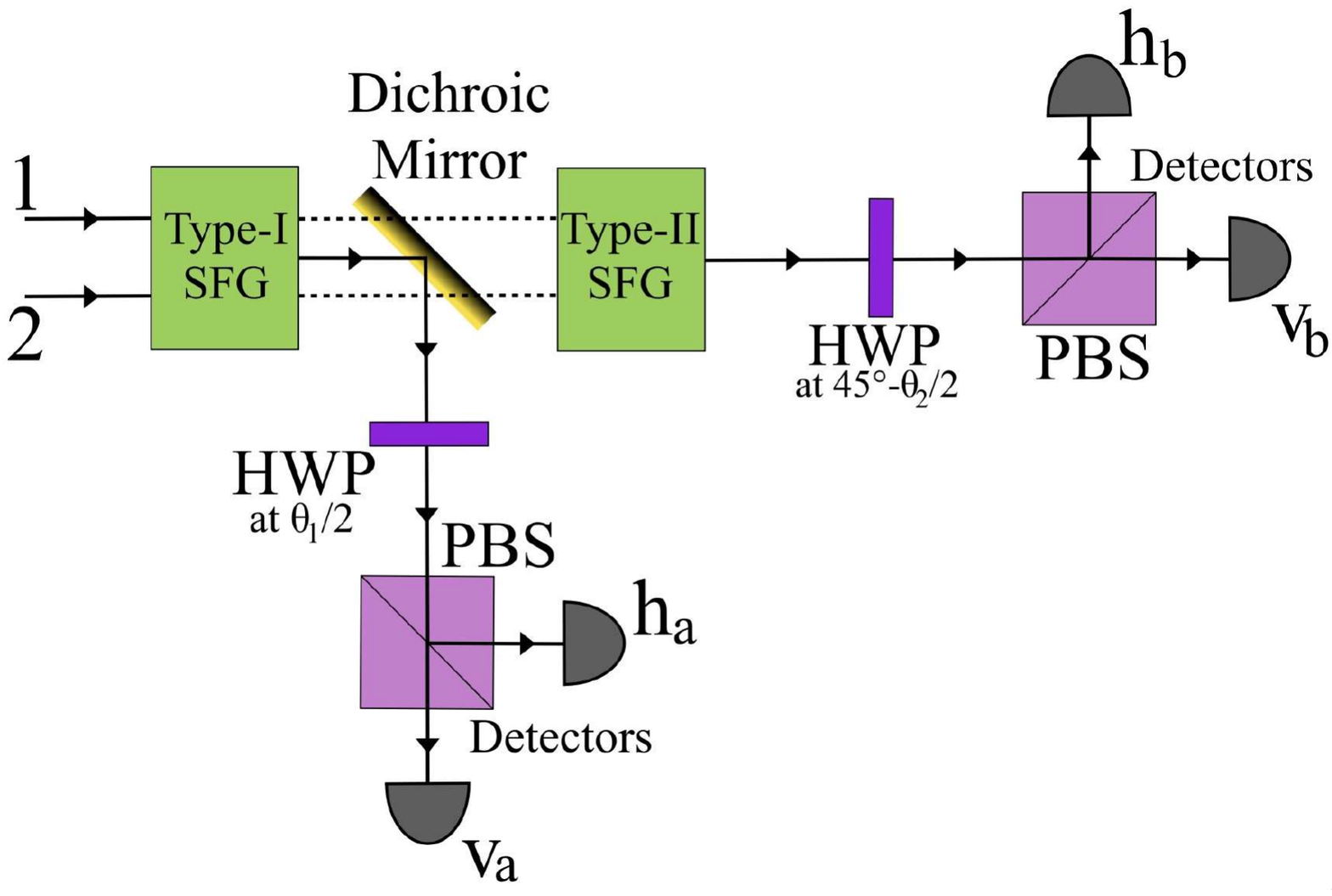}
    \caption{Circuit for the Bell-like state analyzer where non-linear processes like SFG are used to assist. }
    \label{fig:4}
\end{figure}

        A non-linear process known as sum-frequency generation(SFG) is used in \cite{PhysRevLett.86.1370} to implement complete Bell-state analysis and then used to implement quantum teleportation with $100\%$ efficiency. A similar scheme is used here. The circuit looks like Fig.(\ref{fig:4})

The photon pair is first passed through a non-linear crystal where it undergoes type-I SFG. Basically, a pair of photons is converted into a single photon with double the frequency. Mathematically, it enables the following transformation
\begin{align}
    \hat{h}^{\dag}_1\hat{h}^{\dag}_2\longrightarrow\hat{v}^{\dag}_3,\\
    \hat{v}^{\dag}_1\hat{v}^{\dag}_2\longrightarrow\hat{h}^{\dag}_3.
\end{align}
This means it will only act on $\vert \psi_1\rangle$ and $\vert \psi_2\rangle$ and leave the remaining two states unchanged. The transformed states are
 \begin{align}
            &\vert \psi_1\rangle\longrightarrow(\sin(\theta_1)\hat{v}^{\dag}_3+\cos(\theta_1)\hat{h}^{\dag}_3)\ket{0},\\
            &\vert \psi_2\rangle\longrightarrow(\cos(\theta_1)\hat{v}^{\dag}_3-\sin(\theta_1)\hat{h}^{\dag}_3)\ket{0}.
        \end{align}
An important thing to note is that these states have double the frequency of the original photons. Then a dichroic mirror reflects this frequency in direction "a" while all other frequencies are transmitted on path "b". On path "a", the photon is passed through an HWP at an angle $\frac{\theta_1}{2}$. The final transformation looks like
 \begin{align}
            \vert \psi_1\rangle\longrightarrow\hat{h}^{\dag }_{3a}\ket{0},\\
            \vert \psi_2\rangle\longrightarrow\hat{v}^{\dag }_{3a}\ket{0}.
        \end{align}
Similarly in path "b" $\vert \psi_3\rangle$ and $\vert \psi_4\rangle$ undergo type-II SFG which can be mathematically written as
\begin{align}
    \hat{v}^{\dag}_1\hat{h}^{\dag}_2\longrightarrow\hat{v}^{\dag}_3,\\
    \hat{h}^{\dag}_1\hat{v}^{\dag}_2\longrightarrow\hat{h}^{\dag}_3.
\end{align}
Then the beam is allowed to fall on an HWP at an angle $45^\circ-\frac{\theta_3}{2}$. The states thus transform as
 \begin{align}
            \vert \psi_3\rangle\longrightarrow\hat{h}^{\dag }_{3b}\ket{0},\\
            \vert \psi_4\rangle\longrightarrow\hat{v}^{\dag }_{3b}\ket{0}.
        \end{align}
Thus, we have complete Bell-like state discrimination. The signature is:
\begin{itemize}
    \item The state is $\vert \psi_1\rangle$ if detector $h_{a}$ clicks.
     \item The state is $\vert \psi_2\rangle$ if detector $v_{a}$ clicks.
      \item The state is $\vert \psi_3\rangle$ if detector $h_{b}$ clicks.
       \item The state is $\vert \psi_4\rangle$ if detector $v_{b}$ clicks.
\end{itemize}
Thus, we have $100\%$ success probability as compared to a scheme involving hyperentanglement where success probability is only $50\%$. Here we want to point out that the circuit used in \cite{PhysRevLett.86.1370} and above only differ in the orientations of the half-wave plates. Thus, without the use of any extra resource in comparison to the Bell state discrimination, we can distinguish Bell-like states.

All these results can  be summarised in the form of the following table:
\onecolumngrid
   \begin{center}
\begin{tabular} {  | c | c | c | c | c | c | }
    \hline
    \multicolumn{2} { | c | }{\bf Discrimination protocol} & \multicolumn{2} { | c | }{\bf Success Probability} & \multicolumn{2} { | c | }{\bf No. of states discriminated}\\
    \hline
    \multicolumn{2} { | c | }{Using Hyperentanglement} & & & &\\
    \hline
    System qubits & Ancillary qubits & Bell states & Bell-like states & Bell states & Bell-like states \\
    \hline
     Polarisation DOF & Spatial DOF & $100\%$ & $50\%$ & All four & All four\\
     Spatial DOF & Polarisation DOF & $100\%$ & $50\%$ & All four & All four \\
     Polarisation DOF & OAM DOF & $100\%$ & $50\%$ & All four & All four \\
     Polarisation DOF & Time DOF & $100\%$ & $37.5\%$ & All four & All four\\
     \hline
     \multicolumn{2} { | c | }{Using extra ancillary} & $75\%$ & $>25\%$ & All four & Only three (if $\theta_1\neq\theta_2)$ \\
     \multicolumn{2} { | c | }{photon pair} & & (depends on $\theta_1,\theta_2$)&  & All four (if $\theta_1=\theta_2$)\\
     \hline
     \multicolumn{2} { | c | }{Using SFG} & $100\%$ & $100\%$ & All four & All four \\
    \hline
    \end{tabular}
\end{center}
\twocolumngrid
\section{Conclusion and Outlook:}\label{sec8}
In this work, we explored various methods of increasing the success probability of discrimination of Bell-like states using extra resources such as hyperentanglement (entanglement in other DOF of photons), correlation in time, and ancillary entangled qubits in the backdrop of linear optics. Besides that, we have also considered some amount of non-linear resources required to increase this success probability (particularly SFG).

In the case where momentum (or spatial) DOF of photons act as system qubits while polarisation DOF acts as ancillary qubits, we have seen that the success probability of Bell-like state discrimination increases from $25\%$ to $50\%$. Also, all four Bell-like states can be distinguished from each other. Comparing it to the case where we have Bell states instead, we observe that all four Bell states can also be distinguished from one another. The only difference is that it is a perfect discrimination \textit{i.e.}, the success probability is $100\%$. Instead of the above scenario, if we have polarisation DOF as system qubits and momentum DOF as system qubits, we obtain a similar result. The success probability is $50\%$. Using OAM degrees of freedom as ancillary qubits while keeping polarisation DOF fixed as system qubits leads to the same observation. In all three cases, the success probability of unambiguous Bell-like state discrimination becomes $50\%$. Here, it is important to reiterate the point that the results obtained above are yet to be shown to be optimal. The \textit{optimal} fidelity may be higher than $50\%$, but we guess that it is not the case. We try to argue heuristically that the results obtained in this paper indeed might be optimal, but the optimality analysis in general seems very involved.

We have also tried to tackle the problem of Bell-like state discrimination by introducing the correlations in the time DOF of two photons through the use of birefringent materials. Here we obtain an interesting result. In the case of Bell state discrimination, a $100\%$ success probability can be achieved through this method, which automatically implies that all four Bell states can be perfectly distinguished. What we have observed is that in the case of Bell-like states, all four of them can be distinguished from one another but the success probability is $37.5\%$. Another interesting observation arises when we use ancillary qubits to distinguish between Bell-like states. In the Bell state scenario, by adding two ancillary photons, the optimal success probability reaches $75\%$ from $50\%$ and further increases with the number of ancillary photon pairs. However, this is not the case with Bell-like states. Firstly, just by using two ancillary photons, the success probability can be higher or lower than $25\%$ depending on the state parameters $\theta_1$ and $\theta_2$. Secondly, by adding more ancillary photons, instead of increasing, in stark contrast to the Bell states scenario, it decreases. But there is one caveat. The scheme discussed here may not be the optimum situation. As a result, it may still happen that in the case of discrimination of Bell-like states, adding more and more ancilla entangled photon pairs, one may actually increase the success probability. But as mentioned earlier, the optimality analysis for ancilla-assisted state discrimination is quite involved \cite{PhysRevA.98.042323}.

From the above discussion, one can conclude that discrimination of Bell-like states would require more resources as compared to discrimination of Bell states in the linear optical setting. We cannot simply generalize the schemes for Bell-state discrimination to the case of Bell-likestates discrimination. However, we have found that when we use a non-linear element (namely SFG) for Bell-like state discrimination, it can be accomplished in the same amount of resources as required for Bell state discrimination. The only difference, indeed, between both setups is the orientation of the half-wave plates.

It has been shown that in a generalized three-party entanglement-swapping scenario between Alice, Bob, and Charlie where the aim is to create the state having the highest teleportation fidelity between Alice and Charlie,  in some cases, non-maximally entangled measurements (namely Bell-like measurements) outperform Bell measurements or any maximally entangled measurement\cite{quantteleport,ghosal2024repeater}. The experimental realization of such measurements can be done with higher efficiency using the results mentioned in this paper. Also, as mentioned above, the optimality of the given results has still not been analyzed rigorously, which can be another worthwhile direction to explore in order to understand the limitations posed by linear optical schemes in the implementation of various information-theoretical tasks.
\bibliographystyle{apsrev4-2}
\bibliography{ref}

@article{PhysRevA.106.023706,
  title = {Optimal unambiguous discrimination of Bell-like states with linear optics},
  author = {Fields, Dov and Bergou, J\'anos A. and Hillery, Mark and Santra, Siddhartha and Malinovsky, Vladimir S.},
  journal = {Phys. Rev. A},
  volume = {106},
  issue = {2},
  pages = {023706},
  numpages = {8},
  year = {2022},
  month = {Aug},
  publisher = {American Physical Society},
  doi = {10.1103/PhysRevA.106.023706},
  url = {https://link.aps.org/doi/10.1103/PhysRevA.106.023706}
}

@article{PhysRevA.98.042323,
  title = {Ancilla-assisted linear optical Bell measurements and their optimality},
  author = {Olivo, Andrea and Grosshans, Fr\'ed\'eric},
  journal = {Phys. Rev. A},
  volume = {98},
  issue = {4},
  pages = {042323},
  numpages = {10},
  year = {2018},
  month = {Oct},
  publisher = {American Physical Society},
  doi = {10.1103/PhysRevA.98.042323},
  url = {https://link.aps.org/doi/10.1103/PhysRevA.98.042323}
}

@article{calsamiglia2001maximum,
  title={Maximum efficiency of a linear-optical Bell-state analyzer},
  author={Calsamiglia, John and L{\"u}tkenhaus, Norbert},
  journal={Applied Physics B},
  volume={72},
  pages={67--71},
  year={2001},
  url={https://link.springer.com/content/pdf/10.1007/s003400000484.pdf},
  publisher={Springer}
}

@article{PhysRevA.84.042331,
  title = {Arbitrarily complete Bell-state measurement using only linear optical elements},
  author = {Grice, W. P.},
  journal = {Phys. Rev. A},
  volume = {84},
  issue = {4},
  pages = {042331},
  numpages = {6},
  year = {2011},
  month = {Oct},
  publisher = {American Physical Society},
  doi = {10.1103/PhysRevA.84.042331},
  url = {https://link.aps.org/doi/10.1103/PhysRevA.84.042331}
}

@article{PhysRevA.58.R2623,
  title = {Embedded Bell-state analysis},
  author = {Kwiat, Paul G. and Weinfurter, Harald},
  journal = {Phys. Rev. A},
  volume = {58},
  issue = {4},
  pages = {R2623--R2626},
  numpages = {0},
  year = {1998},
  month = {Oct},
  publisher = {American Physical Society},
  doi = {10.1103/PhysRevA.58.R2623},
  url = {https://link.aps.org/doi/10.1103/PhysRevA.58.R2623}
}

@article{PhysRevLett.110.260501,
  title = {Beating the One-Half Limit of Ancilla-Free Linear Optics Bell Measurements},
  author = {Zaidi, Hussain A. and van Loock, Peter},
  journal = {Phys. Rev. Lett.},
  volume = {110},
  issue = {26},
  pages = {260501},
  numpages = {5},
  year = {2013},
  month = {Jun},
  publisher = {American Physical Society},
  doi = {10.1103/PhysRevLett.110.260501},
  url = {https://link.aps.org/doi/10.1103/PhysRevLett.110.260501}
}

@article{PhysRevA.68.042313,
  title = {Hyperentanglement-assisted Bell-state analysis},
  author = {Walborn, S. P. and P\'adua, S. and Monken, C. H.},
  journal = {Phys. Rev. A},
  volume = {68},
  issue = {4},
  pages = {042313},
  numpages = {5},
  year = {2003},
  month = {Oct},
  publisher = {American Physical Society},
  doi = {10.1103/PhysRevA.68.042313},
  url = {https://link.aps.org/doi/10.1103/PhysRevA.68.042313}
}

@article{barreiro2008beating,
  title={Beating the channel capacity limit for linear photonic superdense coding},
  author={Barreiro, Julio T and Wei, Tzu-Chieh and Kwiat, Paul G},
  journal={Nature physics},
  volume={4},
  number={4},
  pages={282--286},
  year={2008},
  url={https://www.nature.com/articles/nphys919},
  publisher={Nature Publishing Group UK London}
}

@article{PhysRevLett.118.050501,
  title = {Superdense Coding over Optical Fiber Links with Complete Bell-State Measurements},
  author = {Williams, Brian P. and Sadlier, Ronald J. and Humble, Travis S.},
  journal = {Phys. Rev. Lett.},
  volume = {118},
  issue = {5},
  pages = {050501},
  numpages = {5},
  year = {2017},
  month = {Feb},
  publisher = {American Physical Society},
  doi = {10.1103/PhysRevLett.118.050501},
  url = {https://link.aps.org/doi/10.1103/PhysRevLett.118.050501}
}

@article{PhysRevLett.86.1370,
  title = {Quantum Teleportation of a Polarization State with a Complete Bell State Measurement},
  author = {Kim, Yoon-Ho and Kulik, Sergei P. and Shih, Yanhua},
  journal = {Phys. Rev. Lett.},
  volume = {86},
  issue = {7},
  pages = {1370--1373},
  numpages = {0},
  year = {2001},
  month = {Feb},
  publisher = {American Physical Society},
  doi = {10.1103/PhysRevLett.86.1370},
  url = {https://link.aps.org/doi/10.1103/PhysRevLett.86.1370}
}

@article{lutkenhaus1999bell,
  title={Bell measurements for teleportation},
  author={L{\"u}tkenhaus, Norbert and Calsamiglia, John and Suominen, K-A},
  journal={Physical Review A},
  volume={59},
  number={5},
  pages={3295},
  year={1999},
  publisher={APS}
}

@article{RevModPhys.79.135,
  title = {Linear optical quantum computing with photonic qubits},
  author = {Kok, Pieter and Munro, W. J. and Nemoto, Kae and Ralph, T. C. and Dowling, Jonathan P. and Milburn, G. J.},
  journal = {Rev. Mod. Phys.},
  volume = {79},
  issue = {1},
  pages = {135--174},
  numpages = {0},
  year = {2007},
  month = {Jan},
  publisher = {American Physical Society},
  doi = {10.1103/RevModPhys.79.135},
  url = {https://link.aps.org/doi/10.1103/RevModPhys.79.135}
}

@article{azuma2015all,
  title={All-photonic quantum repeaters},
  author={Azuma, Koji and Tamaki, Kiyoshi and Lo, Hoi-Kwong},
  journal={Nature communications},
  volume={6},
  number={1},
  pages={6787},
  year={2015},
  publisher={Nature Publishing Group UK London}
}

@article{takeda2019toward,
  title={Toward large-scale fault-tolerant universal photonic quantum computing},
  author={Takeda, Shuntaro and Furusawa, Akira},
  journal={APL Photonics},
  volume={4},
  number={6},
  year={2019},
  publisher={AIP Publishing}
}

@article{knill2001scheme,
  title={A scheme for efficient quantum computation with linear optics},
  author={Knill, Emanuel and Laflamme, Raymond and Milburn, Gerald J},
  journal={nature},
  volume={409},
  number={6816},
  pages={46--52},
  year={2001},
  publisher={Nature Publishing Group UK London}
}

@article{carolan2015universal,
  title={Universal linear optics},
  author={Carolan, Jacques and Harrold, Christopher and Sparrow, Chris and Mart{\'\i}n-L{\'o}pez, Enrique and Russell, Nicholas J and Silverstone, Joshua W and Shadbolt, Peter J and Matsuda, Nobuyuki and Oguma, Manabu and Itoh, Mikitaka and others},
  journal={Science},
  volume={349},
  number={6249},
  pages={711--716},
  year={2015},
  publisher={American Association for the Advancement of Science}
}

@article{reck1994experimental,
  title={Experimental realization of any discrete unitary operator},
  author={Reck, Michael and Zeilinger, Anton and Bernstein, Herbert J and Bertani, Philip},
  journal={Physical review letters},
  volume={73},
  number={1},
  pages={58},
  year={1994},
  url={https://link.aps.org/pdf/10.1103/PhysRevLett.73.58},
  publisher={APS}
}

@article{PhysRevA.69.012302,
  title = {Simple criteria for the implementation of projective measurements with linear optics},
  author = {van Loock, Peter and L\"utkenhaus, Norbert},
  journal = {Phys. Rev. A},
  volume = {69},
  issue = {1},
  pages = {012302},
  numpages = {5},
  year = {2004},
  month = {Jan},
  publisher = {American Physical Society},
  doi = {10.1103/PhysRevA.69.012302},
  url = {https://link.aps.org/doi/10.1103/PhysRevA.69.012302}
}

@article{PhysRevA.73.062320,
  title = {Implementing nonprojective measurements via linear optics: An approach based on optimal quantum-state discrimination},
  author = {van Loock, Peter and Nemoto, Kae and Munro, William J. and Raynal, Philippe and L\"utkenhaus, Norbert},
  journal = {Phys. Rev. A},
  volume = {73},
  issue = {6},
  pages = {062320},
  numpages = {13},
  year = {2006},
  month = {Jun},
  publisher = {American Physical Society},
  doi = {10.1103/PhysRevA.73.062320},
  url = {https://link.aps.org/doi/10.1103/PhysRevA.73.062320}
}

@article{PhysRevA.108.049901,
  title = {Erratum: Method to determine which quantum operations can be realized with linear optics with a constructive implementation recipe [Phys. Rev. A 100, 022301 (2019)]},
  author = {Garcia-Escartin, Juan Carlos and Gimeno, Vicent and Moyano-Fern\'andez, Julio Jos\'e},
  journal = {Phys. Rev. A},
  volume = {108},
  issue = {4},
  pages = {049901},
  numpages = {1},
  year = {2023},
  month = {Oct},
  publisher = {American Physical Society},
  doi = {10.1103/PhysRevA.108.049901},
  url = {https://link.aps.org/doi/10.1103/PhysRevA.108.049901}
}

@article{PhysRevX.8.021017,
  title = {Quantum Optical Realization of Arbitrary Linear Transformations Allowing for Loss and Gain},
  author = {Tischler, N. and Rockstuhl, C. and S\l{}owik, K.},
  journal = {Phys. Rev. X},
  volume = {8},
  issue = {2},
  pages = {021017},
  numpages = {13},
  year = {2018},
  month = {Apr},
  publisher = {American Physical Society},
  doi = {10.1103/PhysRevX.8.021017},
  url = {https://link.aps.org/doi/10.1103/PhysRevX.8.021017}
}

@article{MOYANOFERNANDEZ2017237,
title = {Linear optics only allows every possible quantum operation for one photon or one port},
journal = {Optics Communications},
volume = {382},
pages = {237-240},
year = {2017},
issn = {0030-4018},
doi = {https://doi.org/10.1016/j.optcom.2016.07.085},
url = {https://www.sciencedirect.com/science/article/pii/S0030401816306733},
author = {Julio José Moyano-Fernández and Juan Carlos Garcia-Escartin}
}

@article{williams2017superdense,
  title={Superdense coding over optical fiber links with complete Bell-state measurements},
  author={Williams, Brian P and Sadlier, Ronald J and Humble, Travis S},
  journal={Physical review letters},
  volume={118},
  number={5},
  pages={050501},
  year={2017},
  publisher={APS}
}

@article{zhang2017simultaneous,
  title={Simultaneous entanglement swapping of multiple orbital angular momentum states of light},
  author={Zhang, Yingwen and Agnew, Megan and Roger, Thomas and Roux, Filippus S and Konrad, Thomas and Faccio, Daniele and Leach, Jonathan and Forbes, Andrew},
  journal={Nature communications},
  volume={8},
  number={1},
  pages={632},
  year={2017},
  publisher={Nature Publishing Group UK London}
}

@article{schmid2009quantum,
  title={Quantum teleportation and entanglement swapping with linear optics logic gates},
  author={Schmid, Christian and Kiesel, Nikolai and Weber, Ulrich K and Ursin, Rupert and Zeilinger, Anton and Weinfurter, Harald},
  journal={New Journal of Physics},
  volume={11},
  number={3},
  pages={033008},
  year={2009},
  publisher={IOP Publishing}
}

@article{quantteleport,
  title={Optimal quantum teleportation of collaboration},
  author={Ghosal, Arkaprabha and Ghai, Jatin and Saha, Tanmay and Alimuddin, Mir and Ghosh, Sibasish},
  journal={arXiv preprint arXiv:2401.17201},
  url= {https://arxiv.org/abs/2401.17201},
  year={2024}
}

@article{ghosal2024repeater,
  title={Repeater-Based Quantum Communication Protocol: Maximizing Teleportation Fidelity with Minimal Entanglement},
  author={Ghosal, Arkaprabha and Ghai, Jatin and Saha, Tanmay and Ghosh, Sibasish and Alimuddin, Mir},
  journal={arXiv preprint arXiv:2406.14216},
  url={https://arxiv.org/pdf/2406.14216},
  year={2024}
}

@article{Braun_1995,
  title = {Measurement of the Bell operator and quantum teleportation},
  author = {Braunstein, Samuel L. and Mann, A.},
  journal = {Phys. Rev. A},
  volume = {51},
  issue = {3},
  pages = {R1727(R)--R1730(R)},
  numpages = {0},
  year = {1995},
  month = {Mar},
  publisher = {American Physical Society},
  doi = {10.1103/PhysRevA.51.R1727},
  url = {https://link.aps.org/doi/10.1103/PhysRevA.51.R1727}
}

@article{Weinfurter_1994,
title = {Experimental Bell-State Analysis},
author = {H. Weinfurter},
journal = {Europhysics Letters},
volume = {25},
number = {8},
pages = {559},
month = {mar},
year = {1994},
doi = {10.1209/0295-5075/25/8/001},
url = {https://doi.org/10.1209/0295-5075/25/8/001}
}

@article{bouwmeester1997experimental,
  title={Experimental quantum teleportation},
  author={Bouwmeester, Dik and Pan, Jian-Wei and Mattle, Klaus and Eibl, Manfred and Weinfurter, Harald and Zeilinger, Anton},
  journal={Nature},
  volume={390},
  number={6660},
  pages={575--579},
  year={1997},
  publisher={Nature Publishing Group UK London}
}

@article{Bianchi_2026,
   title={Bell state measurements in quantum optics: a review of recent progress and open challenges},
   volume={11},
   ISSN={2058-9565},
   url={http://dx.doi.org/10.1088/2058-9565/ae609e},
   DOI={10.1088/2058-9565/ae609e},
   number={2},
   journal={Quantum Science and Technology},
   publisher={IOP Publishing},
   author={Bianchi, Luca and Marconi, Carlo and Bacco, Davide},
   year={2026},
   month=Apr, pages={023001} }
\end{document}